\documentclass[aps,pra,10pt,twocolumn,floatfix,superscriptaddress,nofootinbib]{revtex4}

\usepackage[nointlimits]{amsmath}
\usepackage{amssymb, cancel}
\usepackage{exscale}
\usepackage[T1]{fontenc}
\usepackage{graphicx}
\usepackage{color, subfigure}
\usepackage{anyfontsize}

\DeclareMathOperator{\tr}{tr}
\DeclareMathOperator{\Tr}{tr}
\DeclareMathOperator{\rk}{rank}
\DeclareMathOperator{\diag}{diag}
\newcommand{\bra}[1]{\ensuremath{\left\langle{#1}\right\vert}}
\newcommand{\ket}[1]{\ensuremath{\left\vert{#1}\right\rangle}}

\newcommand{\dg}{^{\dagger}}
\newcommand{\trp}{^{\sf T}}

\newcommand{\ma}[1]{\left[\begin{matrix} #1 \end{matrix}\right]}

 \newcommand{\lvec}{{{\lambda}}}

\newcommand{\lvecid}{{{\lambda}}^{0}}
\newcommand{\gs}{\psi_{\rm g}}
\newcommand{\id}{^{\rm 0}}

\newcommand{\erf}[1]{Eq.~(\ref{#1})}

\newcommand{\beq}{\begin{equation}}
\newcommand{\eeq}{\end{equation}}
\newcommand{\bqa}{\begin{eqnarray}}
\newcommand{\eqa}{\end{eqnarray}}
\newcommand{\nn}{\nonumber}
\newcommand{\eg}{\emph{e.g.},~}
\newcommand{\ie}{\emph{i.e.},~}
\newcommand{\etal}{\emph{et al.}~}
\newcommand{\app}[1]{Appendix #1}

\def\Id{\mathbf{1}}
\newcommand{\In}{I^{\otimes n}}

\def\R{\mathbb{R}}

\def\Eq{Eq.~\eqref}
\def\Eqs{Eqs.~\eqref}
\def\Zc{\mathcal{Z}}
\def\Ho{H(\lvec)}
\def\Hk{H_k}
\def\Hj{H_j}
\def\w{{w}} 
\def\la{\langle}
\def\ra{\rangle}
\def\Ha{H_{1}}
\def\Hb{H_{4}}

\definecolor{light-gray}{gray}{0.8}
\makeatletter
\renewcommand*{\@fnsymbol}[1]{\ensuremath{\ifcase#1\or *\or \dagger\or \ddagger\or
   \mathsection\or \mathparagraph\or \|\or **\or \dagger\dagger
   \or \ddagger\ddagger \else\@ctrerr\fi}}
\makeatother

\newcommand{\SNL}{Digital \& Quantum Information Systems,
Sandia National Laboratories, Livermore, CA 94550, USA}
\newcommand{\SJTU}{Joint Institute of UMich-SJTU, Key Laboratory of
  System Control and Information Processing (MOE),
  Shanghai Jiao Tong University, Shanghai, 200240, China}

\begin{document}

\title{Reliability of analog quantum simulation}
\author{Mohan Sarovar}
\email{mnsarov@sandia.gov}
\affiliation{These authors contributed equally to this work}
\affiliation{\SNL}
\author{Jun Zhang}
\email{zhangjun12@sjtu.edu.cn}
\affiliation{These authors contributed equally to this work}
\affiliation{\SJTU}
\author{Lishan Zeng}
\affiliation{\SJTU}

\begin{abstract}
Analog quantum simulators (AQS) will likely be the first nontrivial application of quantum technology for predictive simulation. However, there remain questions regarding the degree of confidence that can be placed in the results of AQS since they do not naturally incorporate error correction. Specifically, how do we know whether an analog simulation of a quantum model will produce predictions that agree with the ideal model in the presence of inevitable imperfections? At the same time there is a widely held expectation that certain quantum simulation questions will be robust to errors and perturbations in the underlying hardware. Resolving these two points of view is a critical step in making the most of this promising technology. In this work we formalize the notion of AQS reliability by determining sensitivity of AQS outputs to underlying parameters, and formulate conditions for robust simulation. Our approach naturally reveals the importance of model symmetries in dictating the robust properties. To demonstrate the approach, we characterize the robust features of a variety of quantum many-body models.
\end{abstract}

\maketitle

Quantum simulation is an idea that has been at the center of quantum
information science since its inception, beginning with Feynman's
vision of simulating physics using quantum
computers~\cite{Feynman:1981tf}.  A quantum simulator is a tunable,
engineered device that maintains quantum coherence among its degrees of freedom over long enough timescales to extract information that is not efficiently computable using classical computers. The modern view of quantum simulation
differentiates between \emph{digital} and \emph{analog} quantum
simulations. Specifically, the former performs simulation of a quantum
model by using discretized evolutions (\ie gates) 
\cite{Walter:2011fh, Kelly:2015jc, Lamata:2016ju} whereas the latter uses a
physical mimic of the model to infer its
properties~\cite{TomiHJohnson:2014dw}. A crucial issue is that while quantum
error correction can be naturally incorporated into digital quantum
simulation, this does not seem to be possible for AQS, which are
essentially special-purpose hardware platforms built to model systems
of interest.  However, digital quantum simulators are extremely
challenging to build, whereas AQS are more feasible in the near future,
with several experimental candidates already under
study~\cite{Kim:2010vz, Simon:2011vs, Trotzky:2012wb, Greif:2013ky, Hart:2015ex}. Thus a critical question for the quantum simulation field is: as AQS become more sophisticated and begin to model systems that are not classically simulable, can one verify or certify the accuracy of results from systems that are inevitably affected by noises and experimental imperfections? \cite{Hauke:2012dq}.

In response to this challenge, we develop a technique for analyzing
the robustness of an AQS to experimental imperfections. We
specialize to AQS that prepare ground or thermal states of quantum
many-body models since these are the most common types of AQS
currently under experimental development. 

\section{Definitions}  
Define a \emph{quantum simulation model}, notated
$(H, O)$, as consisting of a Hamiltonian $H$ and an observable of interest
$O$ (both Hermitian operators). We write a general Hamiltonian in
parameterized form as $H(\lvec) =\sum\nolimits_{k=1}^K \lambda_k \Hk$,
where $\lvec=(\lambda_1, \dots, \lambda_K)\trp$ denotes the vector of
parameters ($\hbar=1$ throughout this paper). $H_k$ are the terms in the Hamiltonian that are individually tunable through the parameters $\lambda_k$. In addition, we decompose the observable into orthogonal projectors representing individual measurement outcomes $O = \sum\nolimits_{m=1}^M \theta_m P_m$ with $P_m P_n = P_m \delta_{mn}$ \footnote{This decomposition of an observable into a set of operators that represent measurement outcomes (or more formally, POVM elements \cite{mikeandike}), is not unique. However, there will be an experimentally relevant decomposition dictated by the experimental apparatus used to probe the AQS. Our results are not dependent on the particular decomposition chosen and for concreteness we work with the decomposition given here.} . 

The goal of an AQS is to produce the probability distribution of a
measurement of $O$ under a thermal state or ground state of a system
governed by $H(\lvecid)$, where $\lvecid$ denotes the ideal, nominal values of the system parameters. That is, to produce the distribution $p_m(\lvecid) = \tr (
P_m \varrho(\lvecid) )$, $m=1$, $\cdots$, $M$, where $\varrho(\lvecid)
= {e^{-\beta H(\lvecid)}}/{\tr e^{-\beta H(\lvecid)} }$, for some
inverse temperature $\beta = 1/k_B T$, if the goal is to predict
thermal properties of the model; or $\varrho(\lvecid) = 
\ket{\gs(\lvecid)}\bra{\gs(\lvecid)}$ with $\ket{\gs(\lvecid)}$ being
the ground state of $H(\lvecid)$, if the goal is to predict ground
state properties. However, due to inevitable environmental
interactions, miscalibration, or control errors, the parameters $\lambda_k$ can
deviate from their nominal values, which can potentially corrupt AQS predictions. We quantify the reliability of an AQS by
the robustness of this probability distribution with respect to the
deviations of $\lvec$ from its ideal value $\lvecid$.

In general, there is no reason to expect that the prepared state $\varrho(\lvec)$ will be
robust to perturbations of $\lvec$. In fact, we know that for
Hamiltonians that possess a quantum critical point, thermal and ground states
can be extremely sensitive to $\lvec$ around that
point~\cite{Zanardi:2007il, Zanardi:2008ih, Invernizzi:2010fe}.
However, reliable AQS does not require robustness of $\varrho(\lvec)$
around $\lvecid$, but only robustness of the probability distribution
of observable outcomes, $\{p_m\}_{m=1}^M$. The fact that this is a
less demanding requirement is the fundamental reason to expect that
some models may be reliably simulated using AQS.

\section{Quantifying AQS robustness} 
To quantify the reliability or robustness of an
AQS, we begin by utilizing the Kullback-Leibler (KL) divergence to measure the
difference between the measurement probability distributions
$p(\lvec)$ and $p(\lvecid)$ \cite{Cov.Tho-1991}:
$D_{\text{KL}}(p(\lvec)||p(\lvecid))
=\sum\nolimits_m p_m(\lvec) \log \frac{p_m(\lvec)}{p_m(\lvecid)}$.
Assuming that the deviation in parameters from the ideal,
$\Delta\lvec=\lvec-\lvecid$, is small, we expand the KL divergence to
second order to obtain
\beq
  \label{eq:19}
 D_{\text{KL}}(p(\lvec)||p(\lvecid))=\frac12 \Delta\lvec\trp F(\lvecid)
  \Delta\lvec + \mathcal{O}(\|\Delta\lvec\|^3). 
\eeq
The positive semidefinite matrix $F$ is the Fisher information matrix (FIM)
for the model, whose elements are given by \cite{Cov.Tho-1991}:
\beq
  \begin{aligned}
F_{ij}(\lvecid)
=&\left.\sum\nolimits_{m=1}^M
\frac{1}{p_m(\lvec)} \frac{\partial p_m(\lvec)}{\partial \lambda_i}
 \frac{\partial p_m(\lvec)}{\partial \lambda_j } \right|_{\lvec =\lvecid}.
  \end{aligned}
  \label{eq:FIM}
\eeq
In Appendices A \& B we describe how to compute the FIM for a quantum simulation model in closed-form, without using numerical approximations to derivatives.
Note that even though we adopt the KL divergence to motivate the FIM,
\u{C}encov's theorem states that the FIM is the unique Riemannian
metric for the space of probability distributions under some mild
conditions \cite{Campbell:1985kj}, and is therefore a general measure of the sensitivity of the parameterized outcome distribution around $\lvecid$. 
 
We first note that if the parameter deviations, $\Delta \lambda$, are Gaussian distributed with zero mean then the expected KL-divergence can be approximated to second-order by the trace of the FIM. This follows from \erf{eq:19}, and the fact that $\frac{1}{M}\sum_{i=1}^M z_i^{\sf T} A z_i$ is an estimate of the trace of $A$ when the elements of $z_i$ are independent, standard normal variables \cite{Avron:2011hg}. However, we are interested in not only obtaining such an average measure of AQS robustness, but also in understanding the factors that determine robustness, or lack thereof, of a particular model. For this purpose we turn to a spectral analysis of the FIM associated with a quantum simulation
model. Consider the set of eigenvalues $\zeta_k$ and eigenvectors $v_k$
of $F$, with $k$ indexing the eigenvalues in descending order. Since
$F$ is a symmetric matrix, we have $F=\sum_{k=1}^K \zeta_k v_k
v_k^\dagger$. Then the simulation error caused by the deviated
parameter $\lambda$ can be approximated to the second order by
$
\sum_{k=1}^K \frac{\zeta_k}{2} \| v_k^\dagger \Delta\lambda\|^2.
$
This error is influenced by two quantities: the magnitude of the eigenvalues, and the overlap of the eigenvectors with the parameter deviation. We can use this structure to quantify the robustness of AQS outputs to the system parameter deviations around the ideal $\lvecid$. 

A quantum simulation model is \emph{trivially robust} to parameter deviations if all $\zeta_k \approx 0$; \ie $F\approx 0$. In the high temperature limit, $\beta \rightarrow 0$, we can show that $F(\lvecid) \rightarrow 0$ at the rate of $\beta^2$ generically and so all models become trivially robust \app{E}. This is expected since the equilibrium state becomes dominated by thermal fluctuations at high temperatures, and observables become insensitive to underlying Hamiltonian parameters. 

A more interesting way a model can be robust is if the FIM possesses only a small number of dominant eigenvalues that are separated by orders of magnitude from other eigenvalues. In this case, only parameter deviations in the directions given by the eigenvectors of dominant eigenvalues affect the simulation results. For instance, if $\zeta_1$ is the dominant eigenvalue, then the \emph{composite parameter deviation} (CPD) $v_1^\dagger \Delta\lambda$ has the major influence on simulation errors. We refer to AQS models that have FIMs with a few dominant eigenvalues separated by orders of magnitude from the rest as \emph{sloppy models}. This terminology is adopted from statistical physics, where it has been recently established that a wide variety of physical models possess properties that are extremely insensitive to a majority of underlying model parameters
, a phenomenon termed \emph{parameter space compression} (PSC) 
\cite{Machta:2013ue, Transtrum:2015hm}.

Model sloppiness is a prerequisite for non-trivial AQS robustness, since  without this property an AQS can only be robust if most or all Hamiltonian parameters can be precisely controlled, an impractical task as quantum simulation models scale in size. In contrast, given a sloppy quantum simulation model, one only has to control and stabilize a few ($\ll K$) influential CPDs. However, model sloppiness alone is not sufficient for AQS robustness since the practicality of controlling these influential CPDs has to be evaluated within the context of the particular AQS experiment at hand, including its control limitations and error model. In this work we aim for a general analysis and do not focus on any particular AQS implementation. Instead, we demonstrate that many quantum simulation models exhibit model sloppiness, the prerequisite for robustness, and how this can help to identify the parameters that must be controlled in order to produce reliable AQS predictions. 

\section{Analyzing the FIM}
A low rank FIM immediately indicates a sloppy model, and since the rank is an analytically accessible quantity, we can use the FIM rank to study model sloppiness beyond numerical simulations. In particular, in this section we discuss two useful methods for bounding the rank of the FIM for a quantum simulation model. 

We begin by rewriting the FIM in a compact form. Define a matrix $V\in
\R^{K \times M}$, whose $km$-th entry is $\frac{\partial p_m(\lvec)}
{\partial \lambda_k}$, and $\Lambda = \diag\{p_1(\lvec)$,
$p_2(\lvec)$, $\cdots$, $p_M(\lvec)\}$.  Then the FIM can be written
as $F = V \Lambda^{-1} V\dg$.  Here we assume that all $p_m(\lvec)$
are non-zero. In the case when some $p_m(\lvec)$ equal $0$, these elements
and the corresponding rows in $V$ should be removed.

This factorized form of the FIM immediately provides a useful bound on its rank. Notice that
the row sum of $V$ is zero, therefore the rank of $V$ is at most $M-1$, which is an upper bound on the rank of $F$. In many physical situations, it is common that the number of distinct measurement outcomes is much less than the number of model parameters, \ie $M\ll K$. In this case, the rank bound of $M-1$ can immediately signal a sloppy model. An example of this that we shall encounter later is a spin-spin correlation function observable, whence $M=2$ and $K$ typically scales with $n$, the number of spins in the model. 

Next we will show that fundamental symmetries of the quantum simluation model
can reduce the rank of the FIM, and further, that symmetries can be used to deduce the structure of the FIM eigenvectors and characterize the influential CPDs. To do this, we define the symmetry group of a quantum simulation model, $G$, as the largest set of symmetries shared by the Hamiltonian and the observable in the model -- \ie the maximal group of space transformations that leave the Hamiltonian and the observable invariant. Let $\{U_g\}_{g\in G}$ be a faithful unitary representation of this symmetry group for the quantum simulation model \footnote{Explicit unitary representations of symmetry groups for several quantum simulation models are presented in \app{H}.}, and suppose $U_g H_k U_g\dg = H_j$ for some $k$, $j$, $g$. Then in \app{C} we show that $\frac{\partial p_m(\lvec)}{\partial \lambda_k} = \frac{\partial p_m(\lvec)}{\partial \lambda_j}$ for all $m$, under ground or thermal states. Therefore, the spatial symmetry of the model leads to identical rows in $V$, and we see an immediate connection between model symmetry and model sloppiness: a high degree of symmetry yields a significant redundancy in the FIM and only a few non-zero eigenvalues. 

This observation suggests a constructive procedure to formulate an upper bound on the rank of FIM based on model symmetries. Specifically, compute the orbit of $H_k$ under the symmetry group for the quantum simulation model; \ie $\{ U_g H_k U_g\dg \vert g \in G\}$, for all $1 \leq k \leq K$. The number of orbits will be the maximum number of distinct rows in the matrix $V$, and therefore provides an upper bound to the rank of the FIM. 

The repeated rows in $V$ resulting from model symmetries also informs us about the structure of the eigenvectors of the FIM, and as a result, the structure of the influential CPDs. Explicitly, the CPD takes the form (see \app{D}):
\beq
v\dg_k \Delta \lvec = \sum_s \mu^{k}_s(\lvecid,\beta) \sum_{l : H_l\in {\rm Orbit}~s} \Delta \lambda_l,
\label{eq:v1_form}
\eeq
where $s$ indexes the unique orbits, and $\mu^k_s$ is a scalar dependent on the orbit, nominal parameter values and temperature. Although the forms of the CPDs are always determined by the eigenvectors of $F$ and therefore by the symmetries of the model, \ie \erf{eq:v1_form}, the coefficients $\mu^k_s(\lvecid, \beta)$ are temperature-dependent and the structure of the CPD can simplify further if these coefficients become alike or approach zero as temperature changes. We will encounter instances of this in the next section.

\begin{figure}[t!]
\centering 
\includegraphics[width=0.9\hsize]{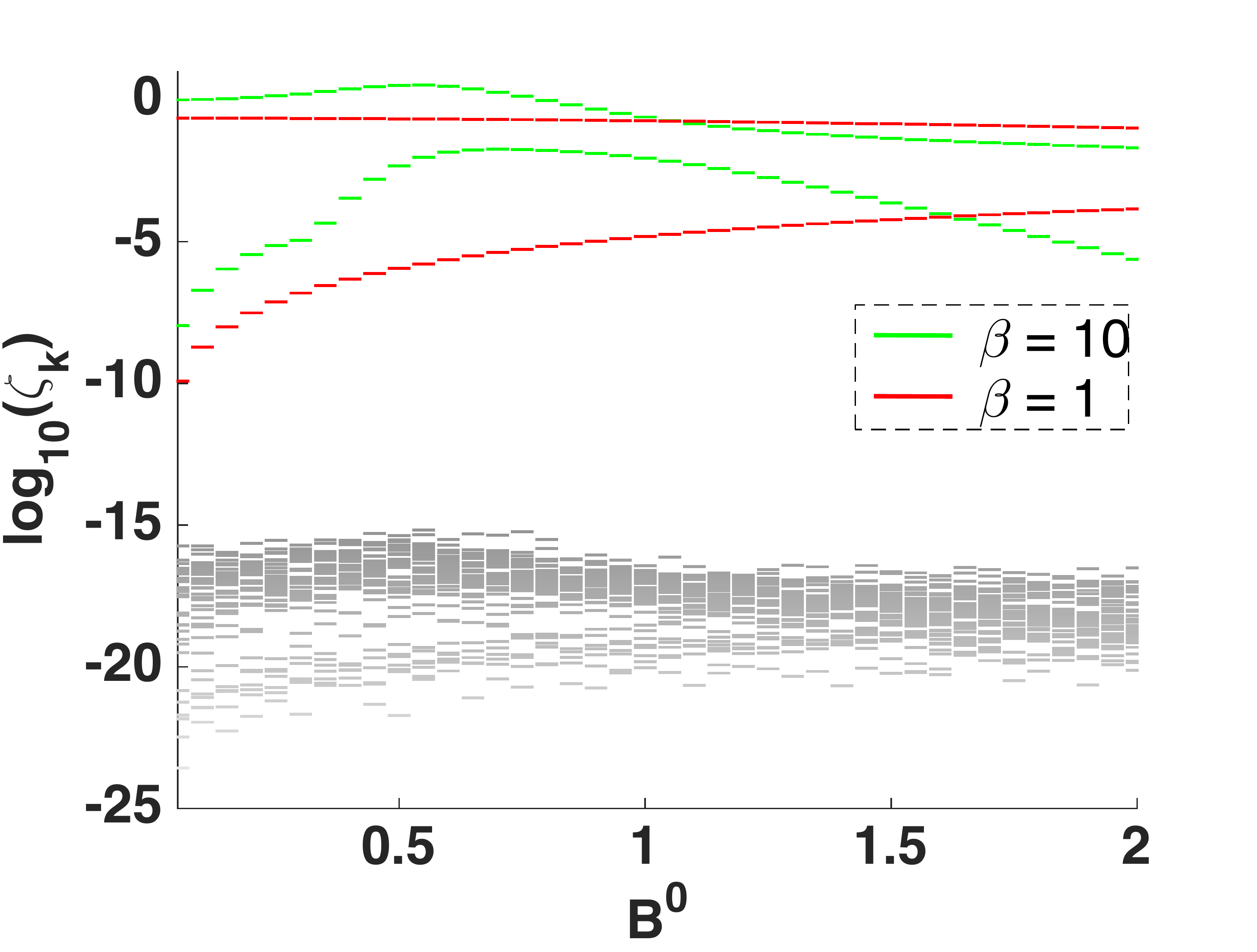}
\caption{Eigenvalues of the FIM for the quantum simulation model $\{\Ha^{\rm per},
  S_z\}$, evaluated for $10$ spins, at low temperature ($\beta=10$)
  and intermediate temperature ($\beta=1$). There are two dominant eigenvalues for all $B^0$ and these are shown in color, while the others are shown in gray.}
\label{fig:1dIsing_unif_Sz_evals}
\end{figure}

\section{Applications}
In this section we use the rank bounds derived above and numerical simulations to understand the sloppiness and robustness of several quantum simulation models. In addition to the applications presented here, we analyze several other quantum simulation models in \app{G}.

\subsection{1D transverse-field Ising model}
The well-known transverse field Ising model in one dimension (1D-TFIM) is described by the Hamiltonian: 
\beq
\Ha = \sum\nolimits_{i=1}^n B_i
\sigma_z^i + \sum\nolimits_{i=1}^n J_{i} \sigma_x^i \sigma_x^{i+1},
\label{eq:1DIsing}
\eeq
where $\sigma_\alpha^i$ is a Pauli operators acting on spin $i$ with
$\alpha=x$, $y$, or $z$, and is normalized such that
$\{\sigma_\alpha, \sigma_\beta\} = \delta_{\alpha\beta} \frac{I}{2}$.
We are interested in the uniform version of this model with
$B^0_i=B^0$ and $J^0_{i}=J^0$ for all $i$; however, when this model is simulated by an AQS, the actual values of $B_i$ and $J_i$ may fluctuate around these nominal values. The boundary conditions for this model can be either periodic, \ie $\sigma_{x}^{n+1} \equiv
\sigma_x^1$, in which case the Hamiltonian will be denoted as
$\Ha^{\rm per}$; or open, \ie $J_n=0$, in which case the Hamiltonian will be denoted as
$\Ha^{\rm open}$.
Although this model is efficiently solvable \cite{Pfeuty:1970wq, Lieb:1961fr, Katsura:1962ja}, its role as a paradigmatic quantum many-body model with a non-trivial phase diagram
makes it a useful benchmark for quantum simulation. Moreover, it exhibits many generic phenomena related to robust AQS, as we will show below.

\begin{figure*}[t!]
\centering
 \subfigure[Eigenvalues of FIM]{ \includegraphics[scale=0.29]{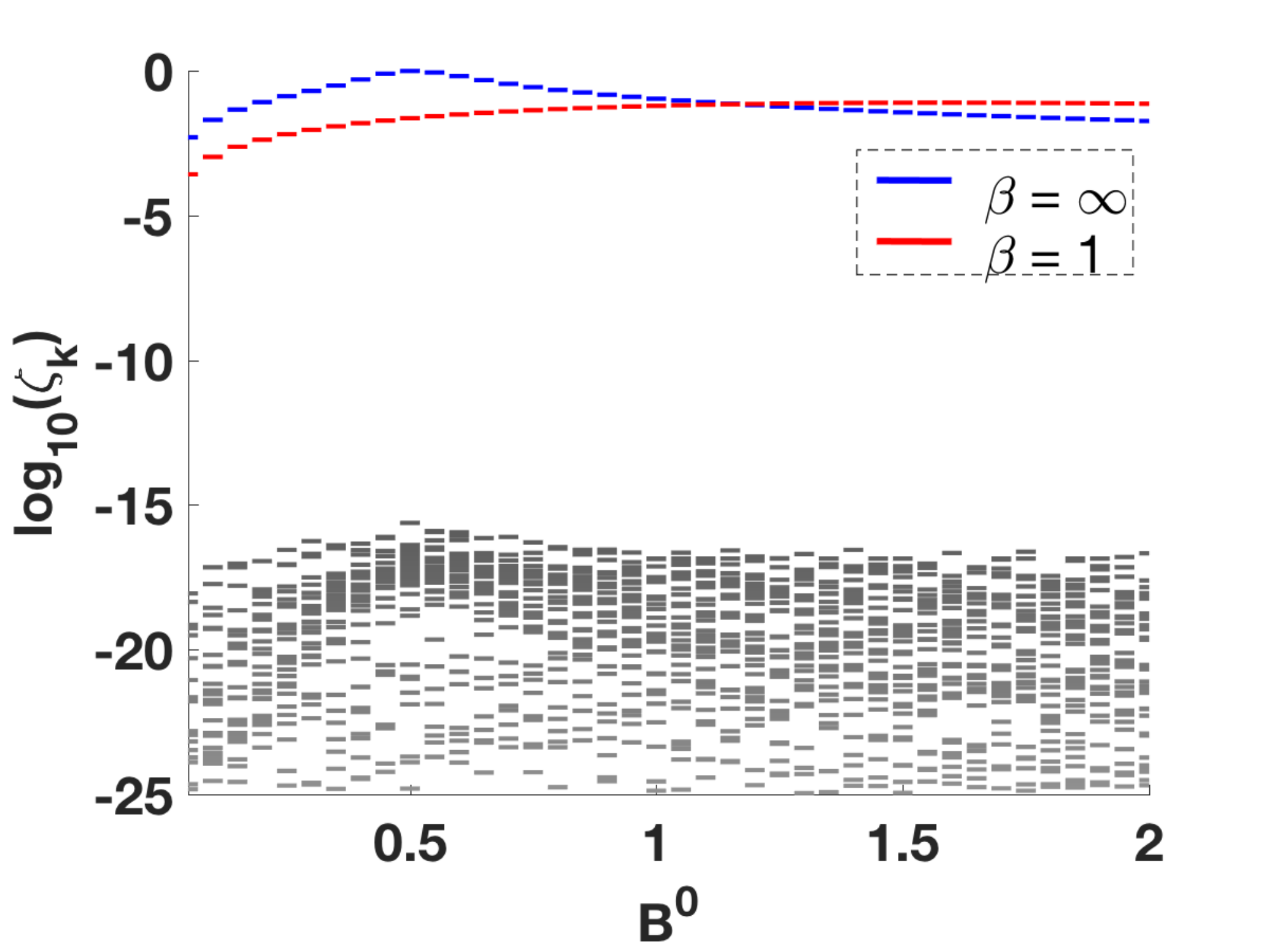} }
 \subfigure[Influential CPD when system is in ground state]{
     \includegraphics[scale=0.15]{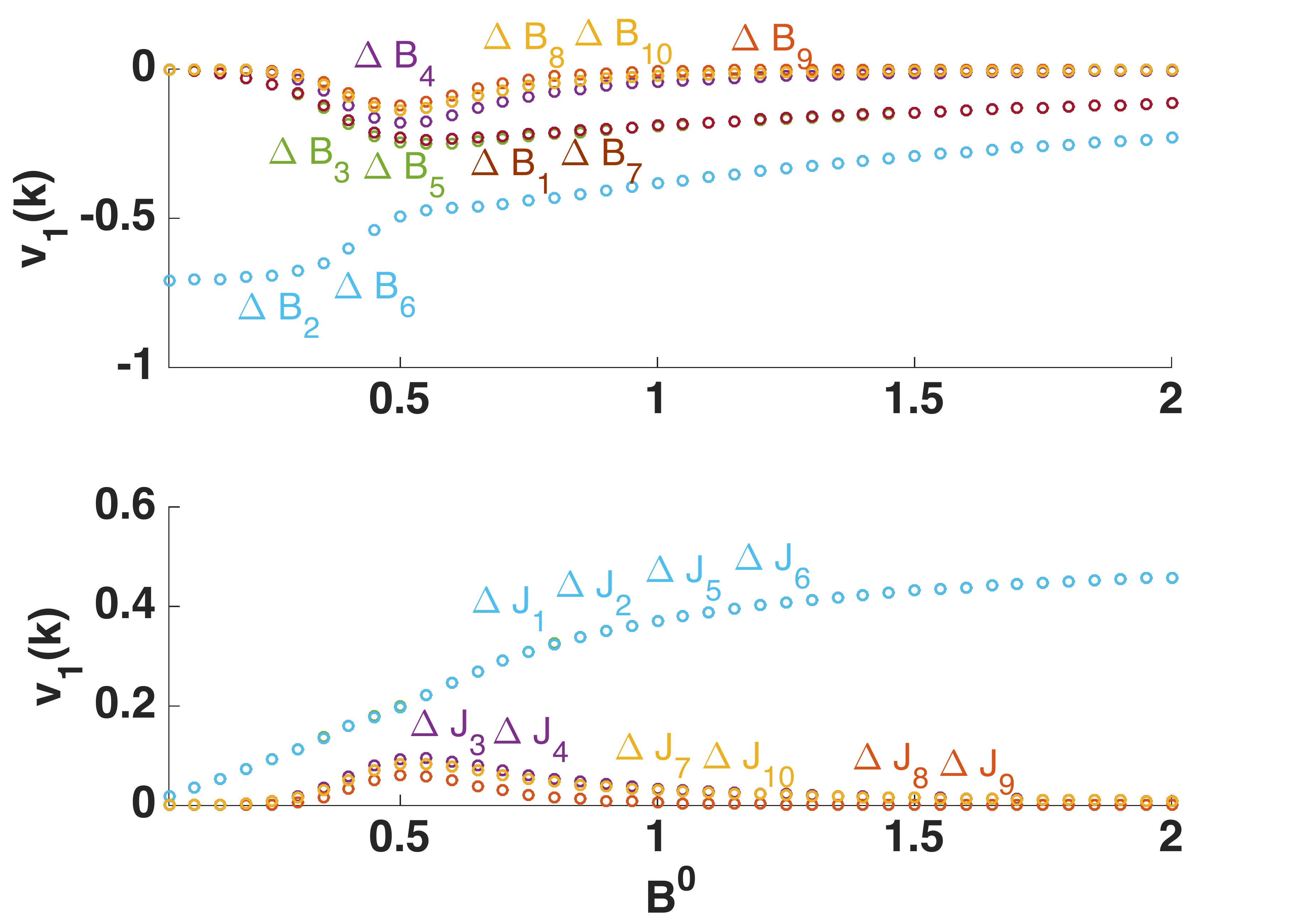} }
 \subfigure[Influential CPD when $\beta=1$]{
     \includegraphics[scale=0.15]{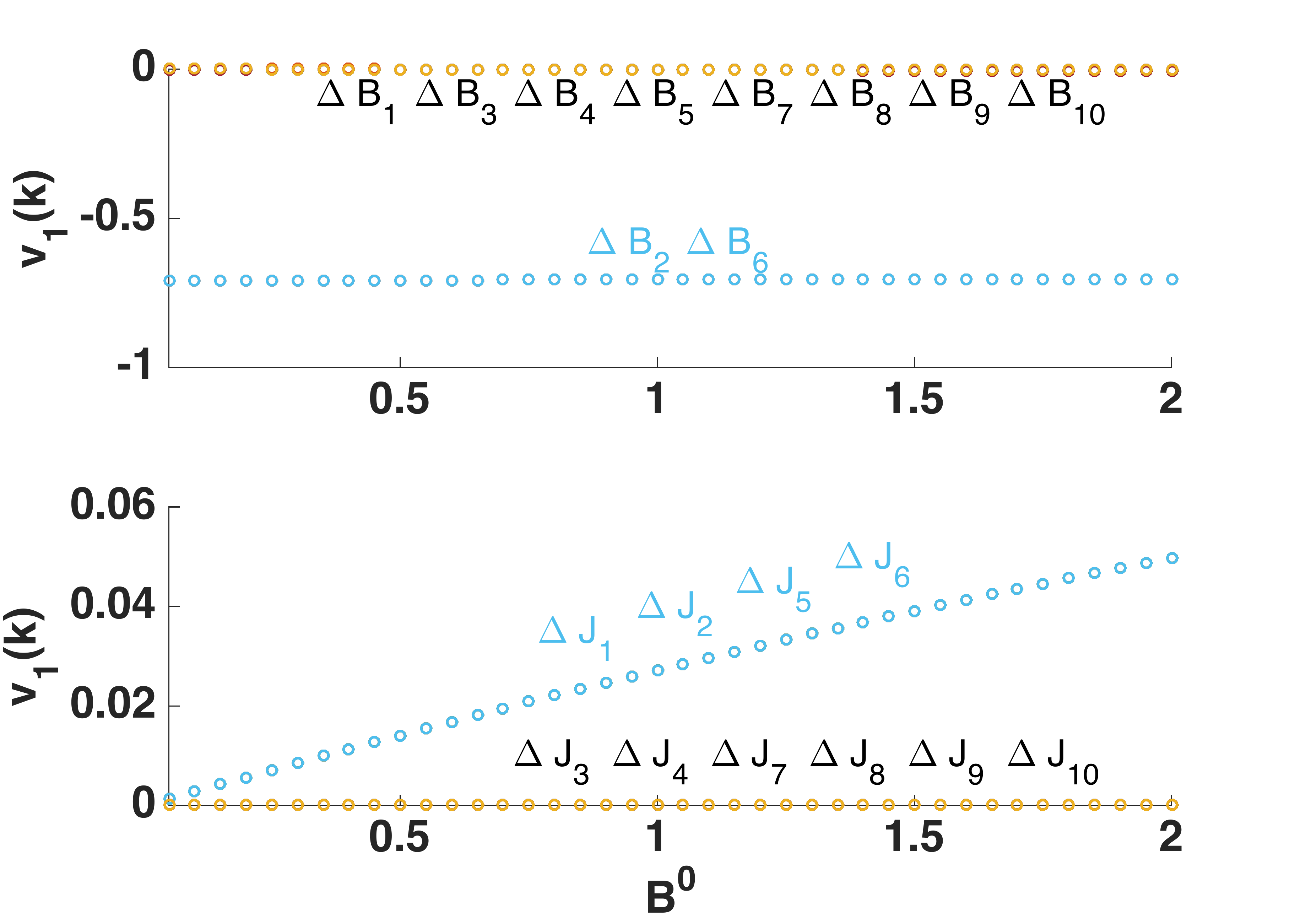} }
\caption{(a) Eigenvalues of the FIM for the AQS model
  $\{\Ha^{\rm per}, C_z(2,6)\}$, evaluated with $10$ spins, at different values of $B^0$. The largest eigenvalue is shown in color for zero temperature (ground state) and intermediate temperature ($\beta=1$), whereas the insignificant ones are shown in gray. (b),(c) Composition of the influential CPD in terms of the original underlying Hamiltonian parameter variations. The data points (which are entries of the principal eigenvector) are labeled by the parameter variation that they multiply to form the CPD, see \erf{eq:v1_form}.}\label{fig:1dIsing_unif_Cx}
\end{figure*}

Two observables of interest in this model are the net transverse magnetization $S_z =
\sum_{i=1}^n \sigma_z^i$ and two-point correlation functions
$C_z(i,j) = \sigma_z^i\sigma_z^j$. It is feasible to measure these
observables experimentally, and importantly, they probe the magnetic
order in the system. For example, both of these observables can be
used to characterize a quantum phase transition that occurs in the
ground state of the uniform 1D-TFIM when swept past its quantum critical point
at $J^0/2B^0=1$ \cite{Chakrabarti:1631381}. 

First we consider the quantum model $\{\Ha^{\rm per}, S_z\}$ with fixed $J\id$, and sweep the parameter $B^0$ to explore the behavior of the model across its phase diagram. This quantum simulation model has full translational invariance. The orbit of any $\sigma_z^i$ under the (lattice) translation group contains all $\sigma_z^j$, $1\leq j \leq n$, and the orbit of any $\sigma_x^i\sigma_x^{i+1}$ contains all $\sigma_x^{j}\sigma_x^{j+1}, ~~ 1 \leq j \leq n$. Consequently, we can prove that 
\beq
\frac{\partial p_m(\lvec)}{\partial B_i} = 
\frac{\partial p_m(\lvec)}{\partial B_j}, ~~ 
\frac{\partial p_m(\lvec)}{\partial J_i} = \frac{\partial p_m(\lvec)}{\partial J_j} 
\eeq
for all $m$ and $1 \leq i, j \leq n$; that is, all the rows in $V$ corresponding to $B$ and $J$ are identical, respectively. Hence, an upper bound on the rank for the FIM of this model is $2$, for all possible $J\id$, $B\id$, $\beta$, and $n$. This is a very sloppy model, especially for large $n$.

To illustrate this general result, in Fig.~\ref{fig:1dIsing_unif_Sz_evals} we show the eigenvalues of the FIM for a $10$-spin 1D-TFIM with $J\id=1$, as $B\id$ is swept. The rank bound derived above is evident in this figure -- there are two dominant eigenvalues -- and the negligible eigenvalues shown in Fig.~\ref{fig:1dIsing_unif_Sz_evals} (gray lines) are actually numerical artifacts. 
In fact, the largest eigenvalue is also orders of magnitude above the second largest, except in the region of the quantum critical point, where the second eigenvalue approaches it (although still many orders of magnitude smaller).

The eigenvectors associated to the two dominant eigenvalues prescribe the parameter deviations that the model is most sensitive to, and due to the full translational invariance of the model we find that they exhibit particularly simple structure (regardless of $\beta$). Namely, the two dominant eigenvectors take the form $[\mu, \cdots \mu, \eta, \cdots, \eta]\trp$ and $[-\eta, \cdots, -\eta, \mu, \cdots, \mu]\trp$, where $\mu$ and $\eta$ are two scalars depending on the value of $B\id$. This implies that across all phases, the model is sensitive only to the CPDs $\sum_{i}\Delta B_i$ and $\sum_i \Delta J_i$. Hence, this quantum simulation model will be robust to parameters deviations as long as these two sums are maintained at zero; \ie local fluctuations of the microscopic parameters that (spatially) average to zero are inconsequential.

Next we examine the AQS model $\{\Ha^{\rm per}, C_z(i,j)\}$ -- \ie
the 1D-TFIM with periodic boundary and a correlation function observable. Noticing that the observable has only two outcomes immediately indicates that the rank of $F$ is at most one, and hence this model is also very sloppy, especially for large $n$. To illustrate this in Fig.~\ref{fig:1dIsing_unif_Cx}(a) we show eigenvalues of the FIM for a $10$-spin example, with the observable being the correlation function $C_z(2,6)$, for zero and intermediate temperature. As expected, only one eigenvalue is significant and all the others are zero up to numerical precision across the whole phase diagram (values of $J^0/2B^0$). 

The structure of the dominant eigenvector is more complex in this case, since although the Hamiltonian is translationally invariant, the observable is not. The eigenvector structure can be extracted from symmetry considerations, but for simplicity we plot its components for the $n=10$ case in Fig.~\ref{fig:1dIsing_unif_Cx}(b), (c), for $\beta=\infty$, $\beta=1$, respectively. Focusing on the zero temperature case first (Fig.~\ref{fig:1dIsing_unif_Cx}(b)), we see that the CPD takes the form $\sum_{i=1}^n \mu_i(B\id) \Delta B_i + \sum_{i=1}^n \eta_i(B\id)\Delta J_i$, where $\mu_i(B\id)$ and $\eta_i(B\id)$ are dependent on $B\id$. Unlike the previous quantum simulation model $\{\Ha^{\rm per}, S_z\}$, the form of the linear combination of underlying model parameters that the AQS is sensitive to not only depends on $B\id$, but this dependence is not the same for all $20$ parameters. Another interesting aspect of Fig.~\ref{fig:1dIsing_unif_Cx}(b) is that away from the quantum critical point, the composite parameter is mostly composed of model parameter variations near the spins whose correlation is being evaluated. More specifically, the AQS model is most sensitive to $(\Delta B_2+\Delta B_6)+(\Delta B_1+\Delta B_3+\Delta B_5+\Delta B_7)/2$ and $(\Delta J_{1}+\Delta J_{2}+\Delta J_{5}+\Delta J_{6})$ (\ie the parameters local to spins involved in the correlation function $C_z(2,6)$). However, near the quantum critical point, all underlying parameter changes enter into the definition of the influential CPD. This is a novel manifestation of collective phenomena in quantum many-body systems: whereas local correlations are typically influenced by local parameters, near a critical point, local correlations are influenced by all the parameters in the system. 

The complexity of the influential CPD for this model is most evident when the system is in its ground state \footnote{This is the reason we present results for the system at zero temperature for this example (instead of $\beta=10$ which is our low temperature case in the other examples).}, but these features persist for small finite temperatures also. However, as shown in Fig.~\ref{fig:1dIsing_unif_Cx}(c), the structure of the CPD simplifies with increased simulation temperature. The sensitivity to all parameter variations in the model around the region near the quantum critical point disappears at intermediate temperature, as expected, since thermal fluctuations overwhelm signatures of quantum criticality as the temperature increases \cite{Cuccoli:2007id}. Moreover, the influential CPD becomes composed of only the parameter changes at the spins involved in the correlation function ($\Delta B_2+\Delta B_6$ and $\Delta J_1+\Delta J_2+\Delta J_5+\Delta J_6$) across the whole phase diagram. 

We pause to reflect on the differences between the two models examined so far. Whereas $\{\Ha^{\rm per}, S_z\}$ and $\{\Ha^{\rm per}, C_z(i,j)\}$ are both sloppy quantum simulation models, the influential CPD for the former is much simpler in form -- its form remains invariant across the phase diagram and with varying temperature. An immediate consequence is that if the goal of a quantum simulation of the 1D-TFIM is to characterize the phase diagram and the phase transition, one should utilize the transverse magnetization as an experimental observable as opposed to correlation functions since the former is more robust to independent local parameter fluctuations. Another option is to probe the site averaged correlation function ($\bar{C}_z(j) = \frac{1}{n}\sum_i \sigma_z^i \sigma_z^{i+j}$) in which case the translational invariance, and consequently robustness to independent local parameter fluctuations of the quantum simulation model is restored. 

\begin{figure}[]
\centering
\subfigure[Eigenvalues of FIM]{
 \includegraphics[width=0.85\hsize]{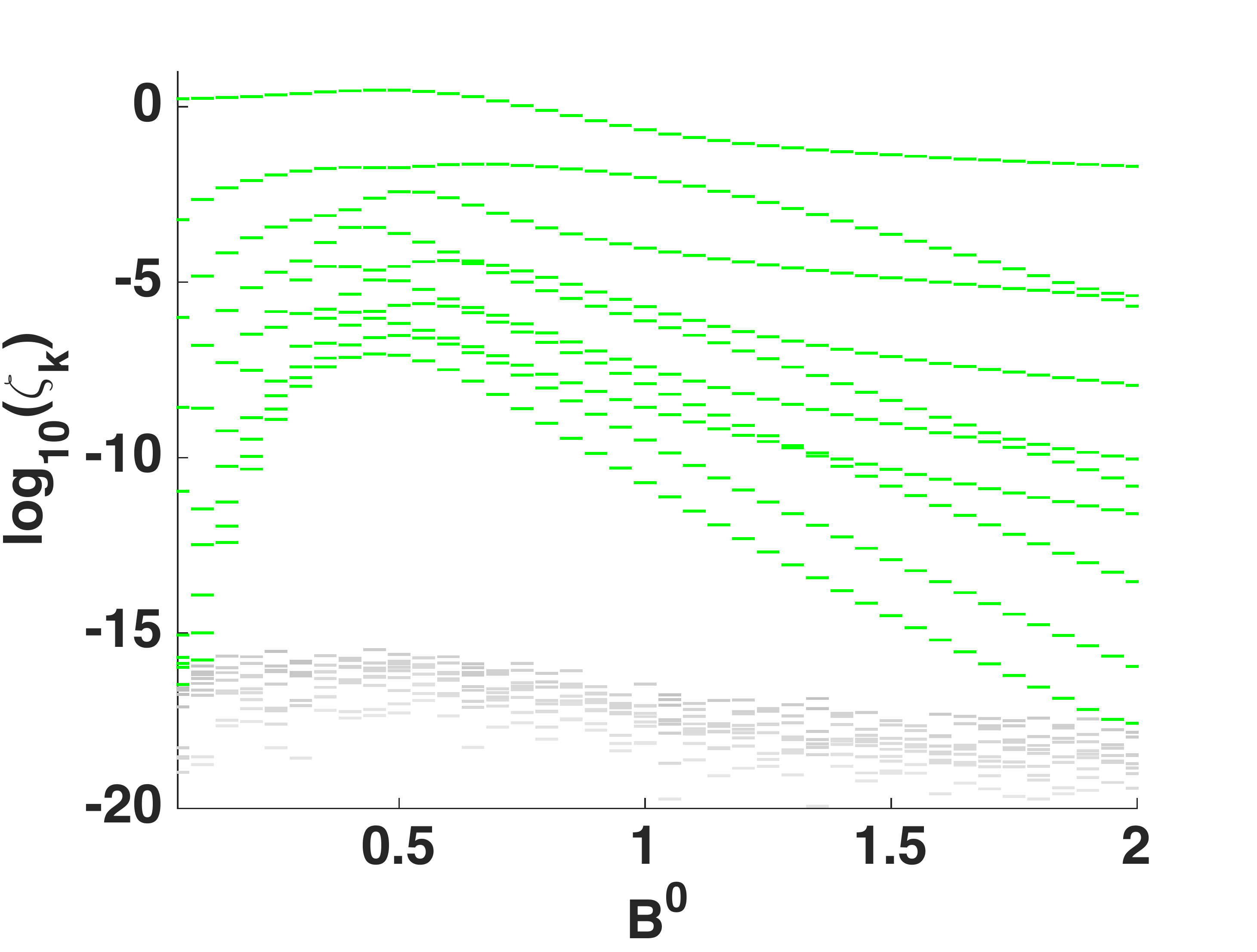} }
\subfigure[Principal influential CPD]{
 \includegraphics[width=0.85\hsize]{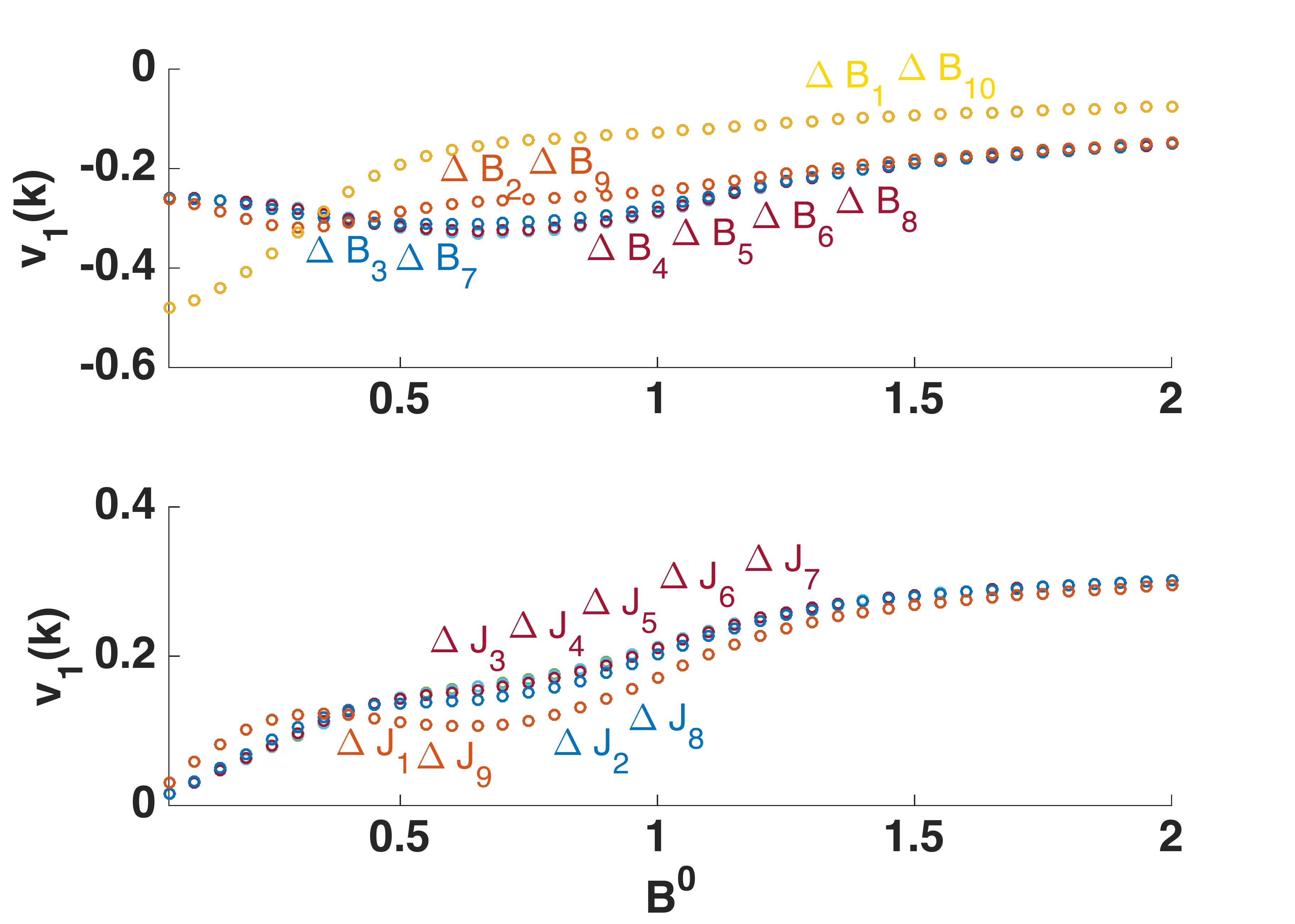} }
\caption{(a) Eigenvalues of the FIM for the quantum simulation model
  $\{\Ha^{\rm open}, S_z\}$, evaluated for a model with $n=10$ spins, at low
  temperature $\beta=10$. The ten largest eigenvalues 
  are shown in red, whereas the others are shown in gray. (b) The
  elements of the eigenvector associated to the largest eigenvalue, which specify the composite influential parameter deviation.}
\label{fig:1dIsing_unif_openbc}
\end{figure}

To study a model with a lower degree of symmetry, we now turn to the
1D-TFIM with open boundary conditions, with the observable of interest being
transverse magnetization again; \ie the quantum simulation model
$\{\Ha^{\rm open}, S_z\}$. This model is no longer translationally invariant,
but has reflection symmetry about the center spin (for odd $n$) or
center coupling (for even $n$). Under this symmetry, each orbit
contains at most two elements -- \eg the orbit of $\sigma_z^j$
contains itself and $\sigma_z^{n+1-j}$ -- and hence an upper bound on the rank of the $(2n-1) \times (2n-1)$ matrix $F$ is $n$. In this case symmetry considerations do not completely reveal the sloppiness of the model, that is, the FIM rank bound is weak, as $n$ is not a lot less than $2n-1$. We explicitly calculate the FIM for this model with $n=10$ at low temperature, and Fig.~\ref{fig:1dIsing_unif_openbc}(a) shows its eigenvalues as a function of $B^0$. As expected from the symmetry rank bound, the model has at most $n=10$ eigenvalues that are nonzero (within numerical precision). Furthermore, the first eigenvalue is several orders of magnitude larger than the others at all phases, although there is a pronounced aggregation of eigenvalues around the quantum critical point. Hence the model is sloppy although not to the same degree as the previous two models examined. The influential CPDs for this model takes the form:
\beq
 \sum_{i=1}^5 \mu_i\left(\Delta B_i+\Delta B_{11-i}\right)
+ \sum_{i=1}^4 \eta_i\left(\Delta J_i +\Delta  J_{10-i}\right) +
\eta_5\Delta J_5, \nn
\eeq
where $\mu_i$ and $\eta_i$ are $B^0$-dependent real numbers. Therefore this model is robust to parameter fluctuations that are negatively correlated across its center spin (or coupling for even $n$). As a result of the complexity of these CPDs and the overall lower degree of sloppiness, we conclude that an AQS implementation of this model will be less robust to parameter fluctuations than the previous two 1D-TFIM models considered.

\subsection{2D transverse field Ising model}
Now we study the uniform 2D-TFIM on an $n\times n$ square lattice:
\beq
H_{2} = \sum\nolimits_{i=1}^{n^2} B_i \sigma_z^i + 
\sum\nolimits_{\langle i,j \rangle} J_{ij} \sigma_x^i \sigma_x^{j}, 
\label{eq:2DIsing}
\eeq
with net magnetization $S_z$ as the observable of interest. Here $\langle
i,j \rangle$ indicates coupling between neighboring spins on a square
lattice. We consider open boundary conditions and the uniform nominal operating point $B_i=B^0$ and $J_{ij}=J^0$. In this case the model has two types of planar symmetries: rotational symmetry about the center of the lattice and mirror reflection symmetry about four reflection lines. The net magnetization observable is invariant under the above symmetries. This is not an exactly solvable model as in the 1D-FTIM case and is therefore of more fundamental interest for AQS.
	
\begin{figure}[t]
\begin{center}
\includegraphics[width=0.4\hsize]{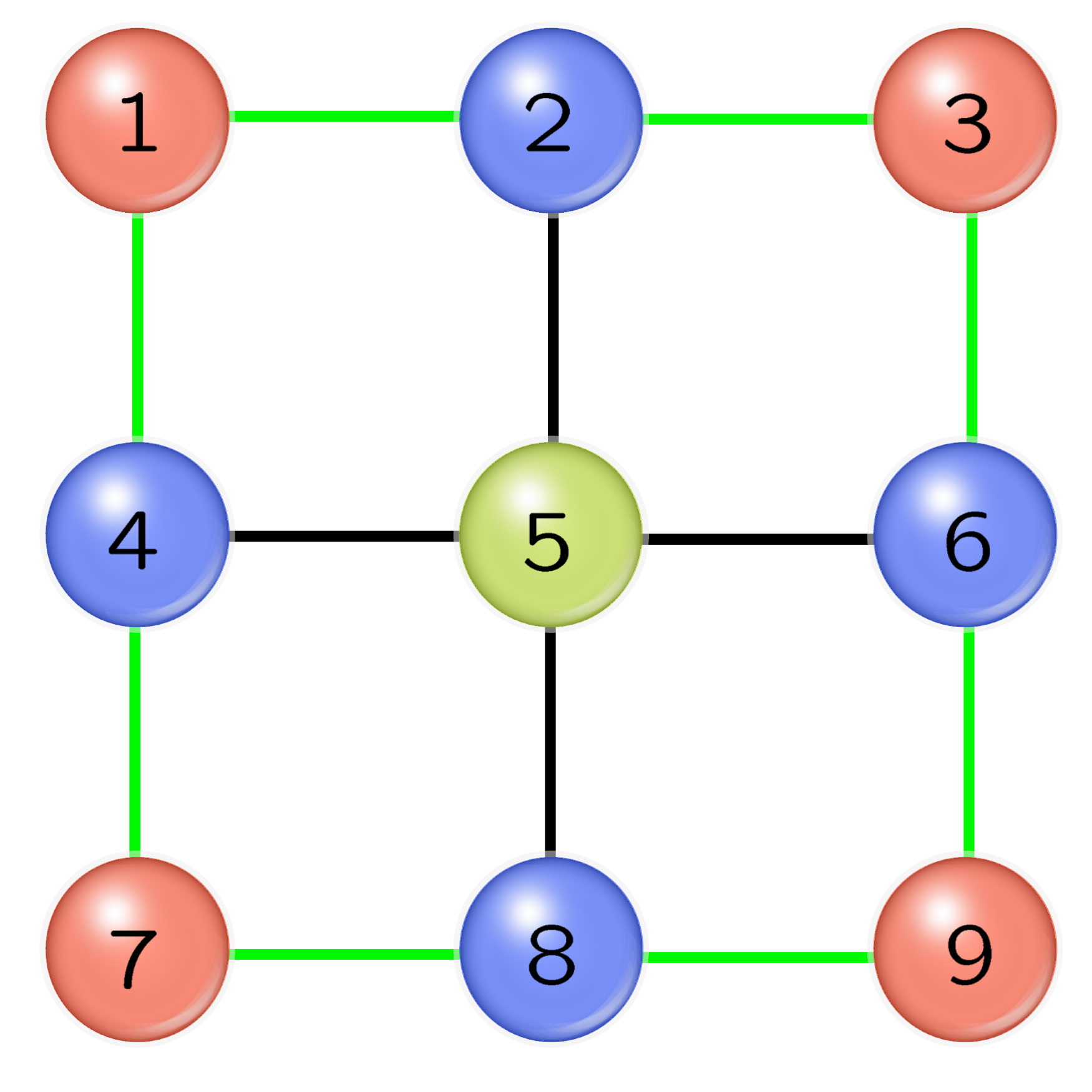}
\end{center}
\caption{Orbits under the symmetry group for the model $\{H_2, S_z\}$ on a $3\times 3$ square lattice. The 2D-TFIM Hamiltonian on this lattice has $\sigma_z$ operators on each site and the links represent the $\sigma_x^i\sigma_x^j$ couplings between sites. Lattice sites and couplings that lie in the same orbit (under the reflections and rotations that leave the quantum simulation model unchanged) are identically colored. There are five distinct orbits in this example.}
\label{fig:2DIsing_symmetries}
\end{figure}

Several local terms ($\sigma_z^i$) and coupling terms ($\sigma_x^i \sigma_x^{j}$) in the Hamiltonian are mapped to the same orbit under the action of the symmetry transformations for $\{H_2, S_z\}$. For example, Fig.~\ref{fig:2DIsing_symmetries} shows the lattice sites and couplings that lie in the same orbit for a $3\times 3$ lattice. There are a total of five distinct orbits in this case and thus the rank the $19\times 19$ FIM is upper bounded by five. Also, according to \erf{eq:v1_form} fluctuations of the local magnetic fields or spin-spin couplings that act on identically colored site or edges in Fig.~\ref{fig:2DIsing_symmetries} will be grouped together in the influential CPD. Explicit computations of eigenvalues and CPDs for this model are included in \app{G.1}.

\subsection{Fermi-Hubbard model}
The Fermi-Hubbard Hamiltonian, a minimal model of interacting electrons in materials, is of significant interest to the AQS community since it is thought that understanding emergent properties of this model could explain some high-$T_c$ superconducting materials \cite{LeBlanc:2015ha}. The Hamiltonian takes the form:
\beq
H_3 = -\sum\nolimits_{\langle i,j \rangle, \sigma} t_{ij}\left( c_{i\sigma}\dg c_{j \sigma} + h.c.\right) + \sum\nolimits_i U_i n_{i\uparrow}n_{i\downarrow},
\eeq
where $c\dg_{i\sigma} (c_{i\sigma})$ creates (annihilates) an electron with spin $\sigma \in \{ \uparrow, \downarrow\}$ on site $i$, $n_{i\sigma} = c\dg_{i\sigma}c_{i\sigma}$ is the electron number operator for site $i$. We consider this Hamiltonian defined over a two-dimensional lattice, and the $\langle i,j \rangle$ indicates that the first sum runs over nearest neighbor sites. Moreover, $t_{ij}$ represents the coupling energy between sites that induces hopping of electrons, and $U_i>0$ represents the repulsive energy between two electrons on the same site. We are interested in the uniform version of this Hamiltonian with nominal parameters $U_i=U^0$, for all $i$ and $t_{ij} = t^0$, for all $i$, $j$. The observable of interest is the double occupancy fraction, $D = \frac{2}{n}\sum_i n_{i\uparrow}n_{i\downarrow}$, where $n$ is the total number of sites, which for example can be used to probe metal to insulator transitions in this model. 

\begin{figure*}[tb]
\centering
 \subfigure[Symmetry of $2\times 3$ Hubbard model]{\raisebox{15mm}{\includegraphics[scale=0.18]{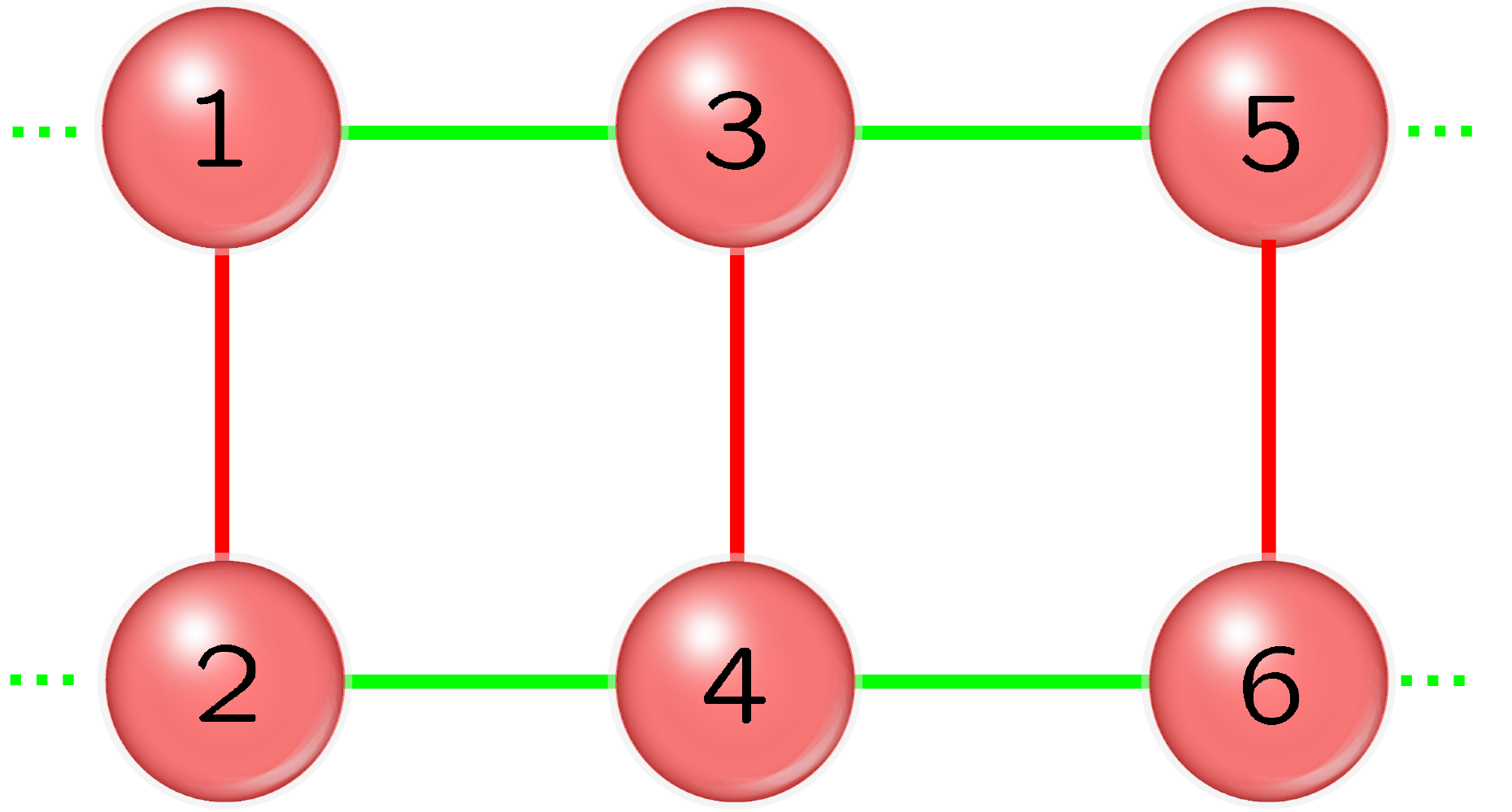}}}
 \subfigure[Eigenvalues of the FIM ]{
     \includegraphics[scale=0.2]{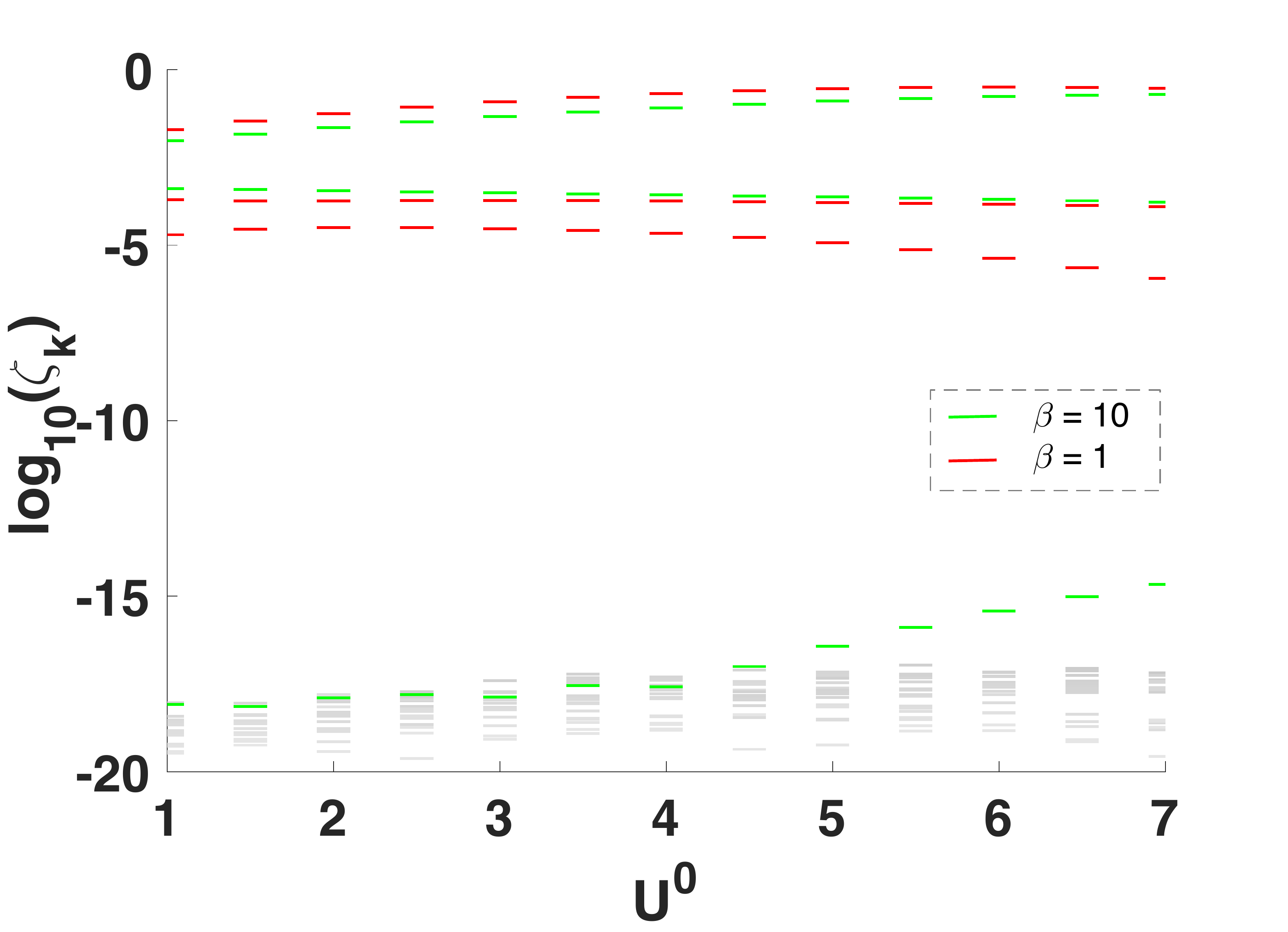} }
 \subfigure[Influential composite parameter deviation, $\beta=10$]{
     \includegraphics[scale=0.2]{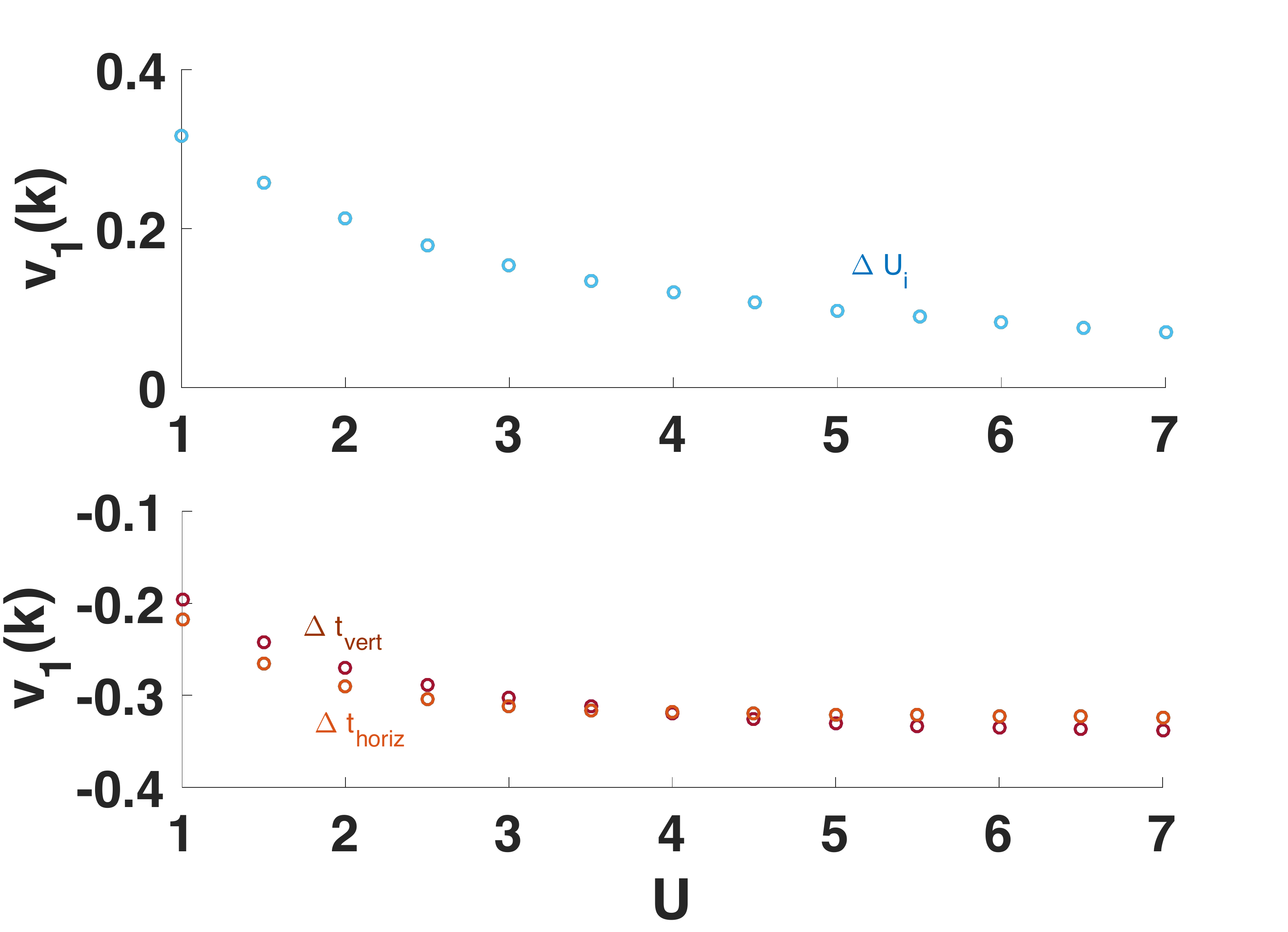} }
\caption{FIM properties for the AQS model $\{H_3, D\}$ at half-filling, for a $2\times 3$ lattice with periodic boundary conditions and $t^0=1$. (a) Orbits under the symmetry operations for this model. The dotted green lines indicate periodic boundary conditions. Lattice sites and couplings that lie in the same orbit are identically colored. There are three distinct orbits in this example. (b) Eigenvalues of the FIM for this model at different values of $U^0$, for $\beta=1$, $10$, with the three largest eigenvalues colored. (c) Composition of the influential CPD in terms of the original underlying Hamiltonian parameter variations, for $\beta=10$. $\Delta t_{\rm vert}$ denotes all the vertical coupling terms (\ie $\Delta t_{12}, \Delta t_{34}, \Delta t_{56}$),and $\Delta t_{\rm horiz}$ denotes all horizontal coupling terms.}
\label{fig:Hubbard}
\end{figure*}

In Fig.~\ref{fig:Hubbard} we show FIM properties for this AQS on a $2\times 3$ lattice with periodic boundary conditions. We show results from simulations of the Hubbard model at half-filling ($\sum_i n_{i\uparrow} = \sum_i n_{i\downarrow} = 3$), but the results are qualitatively the same for the slightly doped cases as well. Fig.~\ref{fig:Hubbard}(a) shows sites and coupling energies that lie within the same orbit under symmetry transformations for this model, which are lattice translations in the $x$ and $y$ directions. All Hamiltonian terms that act locally are mapped between each other and all couplings are mapped between each other, and thus there are three distinct orbits for this model implying an upper bound on the rank of the FIM of 3. Fig.~\ref{fig:Hubbard}(b) shows eigenvalues of the model with $t^0=1$, as a function of $U^0$. As expected, there are always at most three non-zero eigenvalues (to numerical precision) and the model is extremely sloppy. In contrast to the models examined so far, the low temperature version of this model is sloppier than the intermediate temperature version. Finally, Fig.~\ref{fig:Hubbard}(c) confirms that the influential composite parameter deviations take the form expected from the symmetry analysis, with the model only showing sensitivity to the sum of local fluctuations $\sum_i \Delta U_i$, and sum of vertical coupling terms or sum of horizontal coupling terms.

\section{Scaling to large systems}
Quantum simulation is most compelling for large-scale quantum models since difficulty of classical simulation typically increases exponentially with the model scale \footnote{We assume there is some natural notion of scaling of a model that maintains its symmetries -- \eg increasing the number of spins in a spin lattice model while maintaining the coupling configurations.}. Obviously, evaluation of model robustness through classical computation of the FIM is not possible for large-scale models. However, we will show how analysis of small-scale systems can be bootstrapped by various techniques to draw useful conclusions about their large-scale versions. 

First, we note that the bounds on the rank of the FIM that we derived earlier can be useful for models of any scale. For example, the rank bound derived from symmetry considerations allows us to determine the sloppiness of the quantum simulation model $\{\Ha^{\rm per}, S_z\}$ at \emph{any} scale (\ie for any number of spins); and further, symmetry considerations yield the form of the CPD that the model is sensitive to. More generally, we observe that the FIM for any quantum simulation model is greatly simplified by translational invariance, and this can be used to determine sloppiness of the model at any scale. Consider a general (finite-dimensional) translationally invariant Hamiltonian
$H_{g} = \sum\nolimits_{\alpha=1}^{A} \sum\nolimits_\mathcal{N} \lambda^\alpha_{\mathcal{N}} H^\alpha_{\mathcal{N}},
$ 
where $H^\alpha_{\mathcal{N}}$ is an operator acting on degrees of freedom in the spatial neighborhood $\mathcal{N}$, and of type $\alpha$. As an example, consider the following general spin-$1/2$ Hamiltonian on a 3D lattice with nearest-neighbor interactions and periodic boundary conditions in all directions:
\beq
  \begin{aligned}
\Hb&=\sum^n\nolimits_{i=1} \left(B^i_x\sigma^i_x+
B^i_y\sigma^i_y+B^i_z\sigma^i_z\right) \\
&+\sum\nolimits_{\langle i,j \rangle}
\left(J^{ij}_x\sigma_x^i\sigma_x^{j}+
J^{ij}_y\sigma_y^i\sigma_y^{j}+J^{ij}_z\sigma_z^i\sigma_z^{j}\right),    
  \end{aligned}
  \label{eq:genqubitH}
\eeq
where $\langle i,j \rangle$ indicates the sum runs over nearest neighbors in all three directions. Here $\alpha \in \{x,y,z, xx, yy, zz\}$ and the neighborhoods are local sites or edges of the 3D lattice. Translational invariance implies that under the action of the translation symmetry group for these models, all Hamiltonian terms of a given type $\alpha$ lie in the same orbit. Therefore, the number of orbits is the same as the number of types of interaction, and assuming that the observable of interest is also translationally invariant, $A$ is an upper bound on the rank of the FIM for such models at \emph{any} scale. Thus such models are guaranteed to be sloppy, except at very small scales (where the number of parameters is comparable to $A$). Furthermore, the AQS will be most susceptible to the CPDs $\sum \Delta \lambda^\alpha_\mathcal{N}$ for each $\alpha$. For example, for the spin-$1/2$ Hamiltonian $\Hb$ above, if the observable is also translationally invariant, \eg $S_x$, $S_y$ or $S_z$, then the FIM for this quantum simulation model will have rank at most $6$, for \emph{any} number of spins. Note that this example covers a wide range of models including tilted and transverse field Ising models and a variety of Heisenberg models.

The rank bound obtained by counting the number of observable outcomes is also useful in determining sloppiness at any scale. For example, the spin-$1/2$ correlation $C_\alpha(i,j) =\sigma_\alpha^i \sigma_\alpha^j$ has only two possible outcomes $\pm 1$, thus the FIM rank is always one, regardless of the Hamiltonian and number of spins. Unfortunately, this bound does not also inform us about the structure of the CPD that the model is sensitive to. 

\begin{figure}[tb]
\centering
\includegraphics[width=1\hsize]{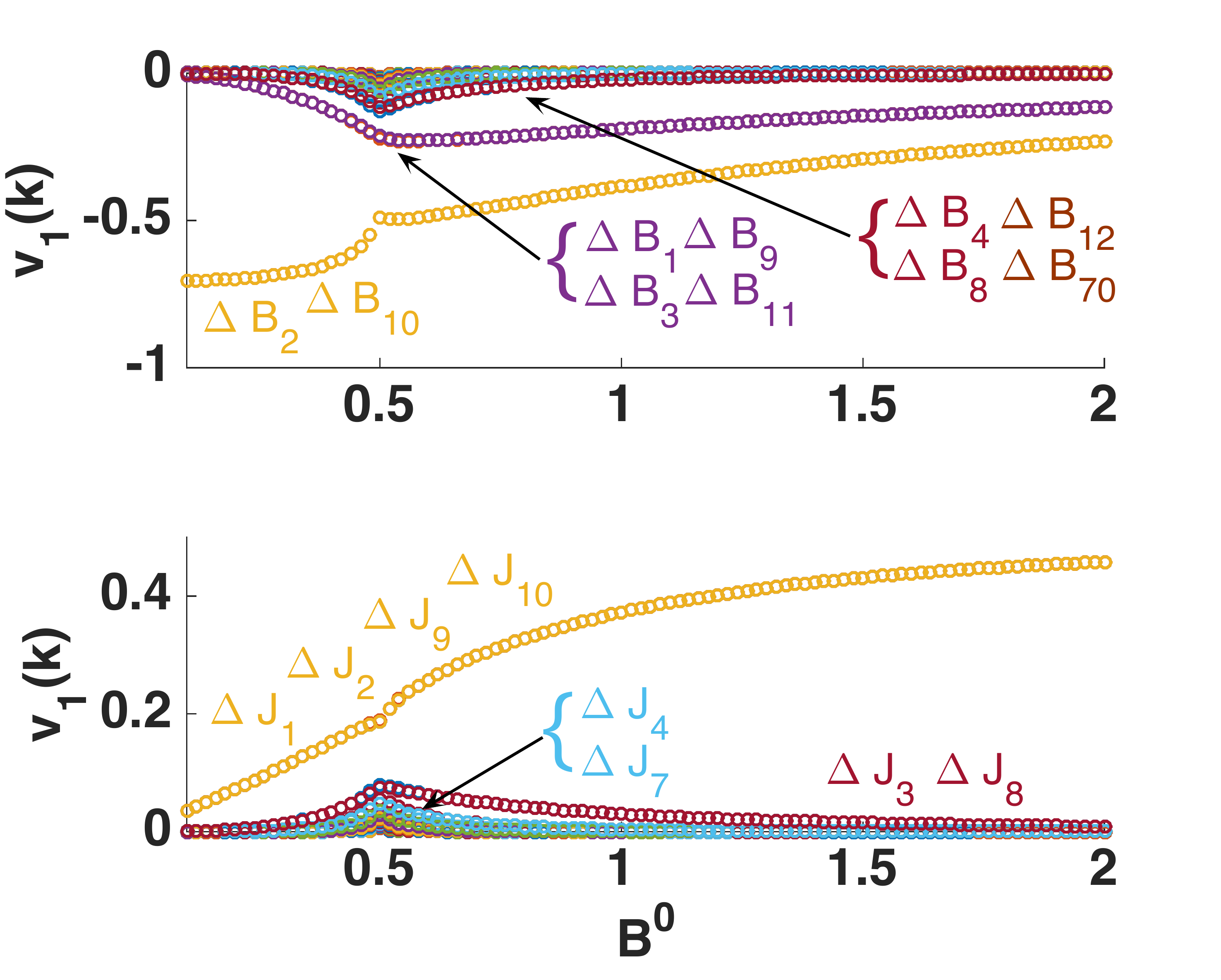}
\caption{Influential CPD for the model $\{\Ha^{\rm per}, C_z(2,10)\}$ evaluated with $n=70$ spins, when the system is in ground state. This model has 140 microscopic parameters, only the ones that significantly contribute to the influential CPD are labeled for clarity.}
\label{fig:5}
\end{figure}

Second, even in cases where a complete symmetry analysis is not possible, an analysis of the small-scale model can be informative about the robustness of the corresponding large-scale model. In particular, since the form of the CPDs is determined by symmetries of the model, one can extrapolate from the form of the CPDs from small-scale models to large versions. For example, for the model $\{\Ha^{\rm per}, C_z(i,j)\}$ studied above, we can examine large-scale behavior by using the well-known exact solution to the 1D-TFIM \cite{Lieb:1961fr, Pfeuty:1970wq} (see \app{F} for details), and confirm that the form of the influential CPD remains the same at large $n$ as for the small-scale version. In Fig.~\ref{fig:5} we plot entries of the dominant eigenvector for the model $\{\Ha^{\rm per}, C_z(2,10)\}$ for $n=70$ spins in the ground state. The influential CPD is mostly composed of parameters around the spins whose correlation function is being evaluated, except near the quantum critical point when other parameters also contribute. These trends agree with results for the small-scale version of the model shown in Fig.~\ref{fig:1dIsing_unif_Cx}(b).

Third, we note that in some cases we can approximate a quantum simulation model with one of higher symmetry in order to gain more information from the FIM. An example of such an approximation is the common practice of imposing periodic boundary conditions on finite lattices in order to make calculations tractable. This approximation can also be useful for assessing robustness of large-scale models using our approach. To illustrate this, we turn to the exact solution of the 1D-TFIM again, and confirm that the model $\{H_1^{\rm open}, S_z\}$ can be approximated by $\{H_1^{\rm per}, S_z\}$ as the number of spins increases. Our numerical investigations show that when $n$ is large, \eg $n>50$, the largest eigenvalue of the FIMs for these two models become almost identical, and the forms of the influential CPDs for the two models approach each other. Hence for some large-scale models one can infer sloppiness and robustness from analysis of approximations with higher degree of symmetry. Of course such approximations are not always possible and one should be aware of their accuracy across parameter regimes.

\begin{figure}[tb]
\centering
\includegraphics[width=0.8\hsize]{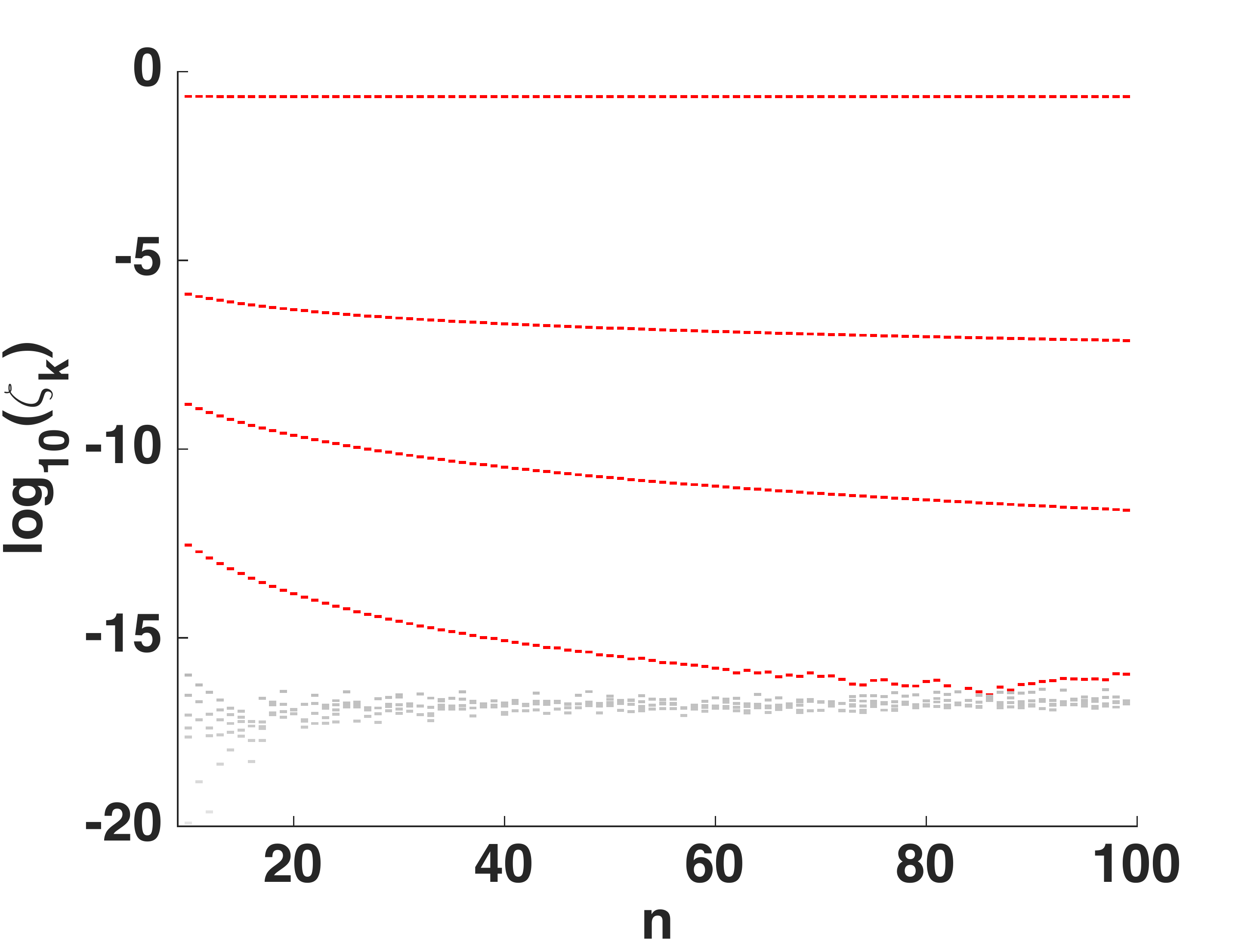}
\caption{The largest 10 eigenvalues of the FIM for the quantum simluation model $\{\Ha^{\rm open}, S_z\}$ as a function of model scale (number of spins, $n$), at intermediate temperature $\beta=1$.}
\label{fig:6}
\end{figure}

Finally, we pose a conjecture regarding the behavior of sloppiness with scale: if a small-scale AQS model with a lattice quantum many-body Hamiltonian is sloppy, then its large-scale version will also be sloppy. Although we currently lack a proof of this statement, it is well supported by numerical evidence.
 For example, consider the model $\{\Ha^{\rm open}, S_z\}$ that was shown to be sloppy at small scales earlier. By utilizing the exact solution to the 1D-TFIM, we can analytically calculate the FIM for a large number of spins. We choose $B\id=0.45$, $J\id=1$, and $\beta=1$, and in Fig.~\ref{fig:6} plot the largest $10$ eigenvalues of the FIM for this model as a function of the number of spins, $n$. The model remains sloppy across all scales that were simulated. 

\section{Discussion}
We have developed and applied a formalism for analyzing the robustness of analog quantum simulators. Many quantum many-body models are potentially robust for AQS, especially if they possess a high degree of symmetry, which we have shown leads to model sloppiness, a necessary condition for robustness. In addition, our techniques allow one to determine which underlying parameter(s) impact simulation results the most, which could help to focus experimental effort when designing AQS platforms. In a sense, our work can be thought of providing a formal justification of the commonly encountered intuition that \emph{bulk properties should be immune to microscopic fluctuations}, and elucidating the connection between this intuition and system symmetries. 

For brevity we have only presented results from applying our approach to uniform models above. However, we have analyzed a large variety of more general models, including ones with random parameters and long-range couplings, and some of the results from these studies are presented in \app{G}. Application of our approach to these more complex cases with less symmetry illustrates how \emph{any} symmetries in the underlying ideal model can be exploited to understand sloppiness and robustness. While nearly all the quantum simulation models we studied were sloppy (the exception being models with complete disorder, \ie random parameters), in some cases the influential CPD is complex, and engineering robust AQS for these models could be challenging. This finding is mirrored by the ubiquity of sloppiness in the classical models studied by Sethna \etal \cite{Machta:2013ue,Transtrum:2015hm}. 

The intent of this work is to introduce the notion of sloppy models to AQS, demonstrate its relation to robust simulation and illustrate that certain quantum simulation models can be robust to uncertainties in parameters. There are many promising directions to extend this work. For example, while we have focused on AQS that prepare ground or thermal states of quantum many-body models, the approach can be extended to analyze quantum simulations that predict dynamic properties of quantum models by considering probability distributions for the dynamical variables of interest. Finally, we have restricted ourselves in this work to investigating the robustness of analog simulation of Hamiltonian models with calibration uncertainties because these uncertainties can in fact dominate the behavior of existing cold-atom analog quantum simulation platforms, \eg \cite{Simon:2011vs, Trotzky:2012wb, Greif:2013ky, Hart:2015ex}, where decoherence due to environmental coupling is very small. However, for a complete picture of robustness, it is desirable to extend this analysis to diagnose robustness of quantum simulation models with decoherence.

\section*{Acknowledgements}
MS would like to acknowledge helpful conversations on the topic of
robust analog quantum simulation with Jonathan Moussa, Ivan Deutsch, Robin
Blume-Kohout, and Kevin Young.  This work was supported by the Laboratory Directed Research and Development program at Sandia National Laboratories. Sandia National
Laboratories is a multiprogram laboratory managed and operated by
Sandia Corporation, a wholly owned subsidiary of Lockheed Martin
Corporation, for the United States Department of Energy's National
Nuclear Security Administration under Contract No. DE-AC04-94AL85000.
JZ thanks the financial support from NSFC under Grant No. 61673264 and 61533012, and State Key Laboratory of Precision Spectroscopy, ECNU, China.

\clearpage
\onecolumngrid
\appendix

\section{Calculation of FIM for thermal states} 
We can analytically
simplify the partial derivatives required to compute the FIM when the
system is in a thermal state $\varrho({\lambda})={e^{-\beta
    \Ho}}/{\mathcal{Z}}$, where $\mathcal{Z}=\tr e^{-\beta \Ho}$. Now
we have
\beq
\label{eq:derivative}
\begin{aligned}
\frac{\partial p_m(\lvec)} {\partial \lambda_k} &=
\frac{\partial }{\partial \lambda_k} \left [
\frac{ \tr (P_m e^{-\beta H(\lvec)})}{\Zc}\right] \\
& =\frac{\tr \left (P_m \frac{\partial e^{-\beta H(\lvec)}} 
{\partial \lambda_k} \right ) \Zc
-\tr(P_m e^{-\beta H(\lvec)})\tr \frac{ \partial e^{-\beta H(\lvec)} }
{\partial \lambda_k} }{ \Zc^2}.
\end{aligned}
\eeq
In order to calculate $ \frac{\partial e^{-\beta H(\lvec)}}{\partial
  \lambda_k}$, we utilize Eq. (78) in Ref. \cite{Najfeld:1995fo} to
obtain:
\beq
\frac{\partial e^{-\beta \Ho}}{\partial \lambda_k} 
= -e^{-\beta H/2} \int_{-\beta/2}^{\beta/2} e^{-\tau H} \Hk e^{\tau H} d\tau
e^{-\beta H/2}.
\label{eq:partial_thermal}
\eeq
Note that we drop the $\lambda$-dependence when it is clear from the
context. Now we diagonalize the Hamiltonian as
\beq
\Ho = T \Gamma T\dg, \nn
\eeq
where $T$ is a unitary matrix of eigenvectors and $\Gamma =
\textrm{diag}\{\gamma_1, \gamma_2, \cdots\}$ is a diagonal matrix of
eigenvalues. Substituting this decomposition into
\erf{eq:partial_thermal}, we get
\beq
\frac{\partial e^{-\beta \Ho}}{\partial \lambda_k} = - T e^{-\beta \Gamma/2} \int_{-\beta/2}^{\beta/2} (T\dg \Hk T)\odot {\Theta}(\tau) ~d\tau\
 e^{-\beta \Gamma/2} T\dg, \nn
\eeq
where $\odot$ denotes the Hadamard product, \ie, element-wise product,
and ${\Theta}_{pq}(\tau)= e^{(\gamma_q-\gamma_p)\tau}$ is the $pq$-th
element of $\Theta$. The $\tau$ dependence is entirely in this matrix, and therefore we can evaluate this integral to yield:
\beq
\frac{\partial e^{-\beta \Ho}}{\partial \lambda_k} = -T e^{-\beta \Gamma/2} ( (T^\dag \Hk T)\odot {\Phi} )
 e^{-\beta \Gamma/2} T\dg, \nn
\eeq
where $\Phi$ is a matrix with elements:
\begin{eqnarray}
 \Phi_{pq}=  \begin{cases}
 \frac{\sinh (\gamma_q-\gamma_p)\beta/2}{(\gamma_q-\gamma_p)/2} ,
&\gamma_p\neq \gamma_q; \\
\beta,&\gamma_p = \gamma_q.
  \end{cases} \nn
\end{eqnarray}
Consequently,
\begin{eqnarray}
  \tr   \frac{ \partial e^{-\beta \Ho} }{\partial \lambda_k}
&=&- \tr   e^{-\beta \Ho} \Hk \beta, \nn \\
\tr  P_m \frac{ \partial e^{-\beta \Ho} }{\partial \lambda_k}
&=&-\tr P_m T e^{-\beta \Gamma/2} ( (T\dg \Hk T)\odot {\Phi} )
 e^{-\beta \Gamma/2} T\dg. \nn
\end{eqnarray}
Inserting these expressions into \erf{eq:derivative} allows us to
evaluate the derivatives required to calculate the FIM for thermal
states in a manner that is numerically stable.

\section{Calculation of FIM for ground states.}
The FIM when the system is in its ground state, $\ket{\gs}$, can also be obtained in an analytical manner. We must calculate
\beq
  \label{eq:25}
\frac{\partial p_m(\lambda)} {\partial \lambda_k} 
=\frac{\partial }{\partial \lambda_k}\tr P_m \varrho_{\rm gs}
=2 \tr P_m \frac{\partial \ket{\gs}}{\partial \lambda_k} \bra{\gs},
\eeq
where $\varrho_{\rm gs} \equiv \ket{\gs}\bra{\gs}$. For a Hamiltonian with a simple (non-degenerate) minimum eigenvalue, the minimum eigenvalue and the associated eigenvector are infinitely differentiable in a neighborhood of $H$, and their differentials at $\Ho$ are \cite{Magnus:85}
\beq
  \label{eq:23}
d E = \bra{\gs} (dH) \ket{\gs}
\eeq
and
\beq
  \label{eq:24}
  d\ket{\gs} = (E_0 I_n-\Ho)^+ (dH) \ket{\gs},
\eeq
where $^+$ denotes the Moore-Penrose (MP) pseudoinverse. We then obtain
\beq
  \label{eq:27}
\frac{\partial \ket{\gs}}{\partial \lambda_k}=  
(E_0 I_n-\Ho)^+ H_k \ket{\gs},
\eeq
and therefore,
\bqa
  \label{eq:derivative_gs}
   \frac{\partial p_m(\lvec)} {\partial \lambda_k}
&=& 2 \tr \left( P_m (E_0 I_n-\Ho)^+ H_k \varrho_{\rm gs} \right)\nn \\
&=& 2 \bra{\gs} P_m (E_0 I_n-\Ho)^+ H_k \ket{\gs}.
\eqa
$V$, the matrix of partial derivatives can then be written in a compact matrix form as:
\beq
  \label{eq:29}
V^{\sf T}=2 \ma{\bra{\gs} P_1 \\ \vdots \\ \bra{\gs} P_{M}}
(E_0 I_n-\Ho)^+ \ma{H_1 \ket{\gs} & \cdots & H_{K} \ket{\gs}}. \nn
\eeq
These analytical expressions for the derivatives for thermal
and ground states are faster and more numerically stable to evaluate than
approximations using difference equations. 

\section{FIM and model symmetries.}  In the main text, we stated that
if a quantum simulation model has a symmetry transformation that
relates $\Hk$ and $\Hj$, then
\beq
  \label{eq:prob_equals}
\frac{\partial p_m(\lvec)} {\partial
  \lambda_k} = \frac{\partial p_m(\lvec)} {\partial \lambda_j}, \quad
\text{for all } m.
\eeq
This has consequences for the rank of the FIM for the model. 

To prove the above, we start with the explicit expressions for the partial derivatives under thermal states, given in \erf{eq:derivative}. The two $k$ dependent quantities in this expression can be written, using \erf{eq:partial_thermal} as:
\beq
  \begin{aligned}
 \tr   \frac{ \partial e^{-\beta H} }{\partial \lambda_k}
&=- \beta\tr   e^{-\beta H} \Hk , \\
\tr  P_m \frac{ \partial e^{-\beta H} }{\partial \lambda_k}
&=-\tr \left( P_m e^{-\beta H/2} \int_{-\beta/2}^{\beta/2} e^{-\tau H} \Hk e^{\tau H} d\tau
e^{-\beta H/2}\right). \nn
  \end{aligned}
\eeq
Then suppose the quantum simulation possesses a symmetry with unitary
representation (we assume the symmetry group is compact) $\{U_g\}_g$,
in which case $[U_g,H(\lvec)] = [U_g,O] = 0$ for all $g$. Furthermore,
given the decomposition of the observable, $[U_g, P_m] = 0, ~~ \forall
g,m$. Now, suppose the symmetry maps $H_j$ to $H_k$, meaning $H_k =
U_g H_j U_g\dg$, then using the commutation properties stated above,
\beq
  \begin{aligned}
 \tr   \frac{ \partial e^{-\beta H} }{\partial \lambda_k}
=- \beta\tr   e^{-\beta H} U_g \Hj U_g\dg 
= \tr   \frac{ \partial e^{-\beta H} }{\partial \lambda_j}. \nn
  \end{aligned}
\eeq
Also, 
\beq
  \begin{aligned}
& \tr  P_m \frac{ \partial e^{-\beta H} }{\partial \lambda_k}\\
=&-\tr \left( P_m e^{-\beta H/2} \int_{-\beta/2}^{\beta/2} 
e^{-\tau H} U_g \Hj U_g\dg e^{\tau H} d\tau e^{-\beta H/2}\right) \\
=&\tr  P_m \frac{ \partial e^{-\beta \Ho} }{\partial \lambda_j}. \nn
  \end{aligned}
\eeq
Therefore, all $k$-dependent terms in \erf{eq:derivative} are the same if we exchange $k$ with $j$, and hence we arrive at \erf{eq:prob_equals} for thermal states.

To prove the same property when the system is in its ground state, we turn to the expression for the partial derivatives given in \erf{eq:derivative_gs}:
\beq
  \begin{aligned}
 \frac{\partial p_m(\lvec)} {\partial \lambda_k}
=&2  \Tr \left( P_m (E_0 I_n-\Ho)^+ U_g H_j
U_g\dg\varrho_{\rm gs}\right),
  \end{aligned}
\eeq
Since $[U_g,\Ho]=0$, and both of these operators are normal, they share an eigenbasis, implying $[U_g, \varrho_{\rm gs}]=0$.  Therefore,
\beq
 \frac{\partial p_m(\lvec)} {\partial \lambda_k}
= 2  \Tr \left( P_m
U_g\dg (E_0 I_n-\Ho)^+ U_g H_j \varrho_{\rm gs}\right).
\label{eq:gs_eq}
\eeq
Using $[U_g,\Ho]=0$, it is easy to verify that $U_g(E_0-\Ho)^+U_g\dg$ is also the MP pseudoinverse of $E_0 I-\Ho$, and from the uniqueness of
MP pseudoinverse, we have that
\beq
  \label{eq:97}
  U_g\dg(E_0 I_n-\Ho)^+ U_g=(E_0 I_n-\Ho)^+.
\eeq
From this equality and \erf{eq:gs_eq}, \erf{eq:prob_equals} follows for ground states as well.

\section{Structure of the eigenvectors of $F$}

As discussed in the main text, spatial symmetries of a quantum simulation model render some rows of the matrix $V$ equal. Here we show that this induces a certain structure on the Fisher information matrix (FIM), namely that the corresponding entries of each eigenvector of $F$ are equal.

Without loss of generality, we assume that $V$ can be written as
\begin{equation}
  \label{eq:2}
V=\ma{\Id_1 v_1^{\sf T} \\ \Id_2 v_2^{\sf T} \\ \vdots \\ \Id_s v_s^{\sf T}},
\end{equation}
where $\Id_k$ is a column vector with dimension $n_k$ and all entries being 1, and
$v_k^{\sf T}$ are pairwise distinct row vectors. As a result,
\begin{equation}
  \label{eq:6}
F=V \Lambda^{-1} V^\dagger
=\ma{ v_1^{\sf T}\Lambda^{-1}v_1 \Id_1 \Id_1^{\sf T} & \cdots 
& v_1^{\sf T}\Lambda^{-1}v_s \Id_1 \Id_s^{\sf T}\\ \vdots & & \vdots \\
v_s^{\sf T}\Lambda^{-1}v_1 \Id_1 \Id_1^{\sf T} & \cdots 
& v_s^{\sf T}\Lambda^{-1}v_s \Id_1 \Id_s^{\sf T}}.
\end{equation}
Let
\begin{equation}
  \label{eq:7}
M=\ma{ v_1^{\sf T}\Lambda^{-1}v_1 & \cdots & v_1^{\sf T}\Lambda^{-1}v_s \\ \vdots & & \vdots \\
v_s^{\sf T}\Lambda^{-1}v_1 & \cdots & v_s^{\sf T}\Lambda^{-1}v_s},
\quad D=\diag\{n_1, \cdots, n_s \},
\end{equation}
and $p^{\sf T}=\ma{p_1 & \cdots & p_s}$ is an eigenvector of $M D$ with
eigenvalue $\alpha$. Then
\begin{equation}
  \label{eq:1}
F\ma{p_1\Id_1 \\ p_2\Id_2 \\ \vdots \\ p_s\Id_s}  
=\ma{v_1^{\sf T} \Lambda^{-1} v_1 p_1 n_1 \Id_1 + \cdots 
+v_1^{\sf T} \Lambda^{-1} v_s p_s n_s \Id_1 \\ \vdots \\
v_s^{\sf T} \Lambda^{-1} v_1 p_1 n_1 \Id_s + \cdots  
+v_s^{\sf T} \Lambda^{-1} v_s p_s n_2 \Id_s }
=\alpha \ma{p_1\Id_1 \\ p_2\Id_2 \\ \vdots \\ p_s\Id_s}.
\end{equation}
Therefore, $\ma{p_1\Id_1^{\sf T} &p_2\Id_2^{\sf T} & \cdots &
  p_s\Id_s^{\sf T}}^{\sf T}$ is an eigenvector of $F$. From \Eq{eq:2},
we know that the rank of $V$ is $s$, and thus the ranks of $M$ and $F$
are both $s$. Hence, all the eigenvectors of $F$ can be written
in the form $\ma{p_1\Id_1^{\sf T} &p_2\Id_2^{\sf T} & \cdots &
  p_s\Id_s^{\sf T}}^{\sf T}$, that is, they have the same structure of
repeated entries as $V$ in \Eq{eq:2}.

\section{Robustness at high temperature} 

We will show that in the limit of high temperature, the FIM approaches
$0$ at the rate of $\beta^2$. For simplicity, we consider an $n$-qubit system.
From \app{A}, and therefore we know that when the
system is in a thermal state $\varrho({\lambda})={e^{-\beta
    \Ho}}/{\mathcal{Z}}$, we have
\begin{equation}
\label{eq:derivative}
\frac{\partial p_m} {\partial \lambda_k}
=\tr \left (P_m \frac{\partial e^{-\beta H}} 
{\partial \lambda_k} \right ) \bigg/ \Zc
-\tr(P_m e^{-\beta H})\tr \frac{ \partial e^{-\beta H} }
{\partial \lambda_k} \bigg/\Zc^2,
\end{equation}
where $\mathcal{Z}=\tr e^{-\beta \Ho}$. 
In the high temperature limit, $\beta\rightarrow 0$, we expand to the
first order
\begin{equation}
  e^{-\beta H} \approx I-\beta H
\end{equation}
to obtain
\begin{equation}
\begin{aligned}
\frac{\partial e^{-\beta H}}{\partial \lambda_k}& \approx -\beta H_k
\end{aligned}
\end{equation}
and
\begin{equation}
\mathcal{Z}=2^n-\beta\tr H,\quad 
\mathcal{Z}^{-1}=2^{-n}+2^{-2n}\beta\tr H,\quad
\mathcal{Z}^{-2}=2^{-2n}+ 2^{-3n+1} \beta \tr H.
\end{equation}
Further, using this approximation and ignoring higher order terms in
$\beta$, we get
\begin{equation}
  \begin{aligned}
\tr \left (P_m \frac{\partial e^{-\beta H}} 
{\partial \lambda_k} \right ) \bigg/ \Zc  
&\approx -\beta \tr P_m H_k (2^{-n}+2^{-2n}\beta\tr H)\\
&\approx - 2^{-n}\beta \tr P_mH_k,                                                                                                                                                                                                                                                                                                                                                                                                                                                                                                                                                             
  \end{aligned}
\end{equation}
and
\begin{equation}
  \begin{aligned}
\tr(P_m e^{-\beta H})\tr \frac{ \partial e^{-\beta H} }
{\partial \lambda_k} \bigg/\Zc^2
&\approx-\beta \tr P_m (I-\beta H) \tr H_k (2^{-2n}+ 2^{-3n+1} \beta\tr H)\\
&\approx-2^{-2n} \beta \tr H_k \tr P_m.
  \end{aligned}
\end{equation}
Combining these two equations, we have
\begin{equation}
\frac{\partial p_m} {\partial \lambda_k} \approx \beta u_{km},
\end{equation}
where
\begin{equation}
  \label{eq:11}
u_{km}=2^{-2n}\tr H_k \tr P_m-2^{-n} \tr P_mH_k.  
\end{equation}
Define a
matrix $U$ whose $km$-th element is $u_{km}$. Then $F=\beta^2
U\Lambda^{-1}U\dg$. Hence, as $\beta \rightarrow 0$, the FIM approaches the
zero matrix as $\beta^2$ and thus the quantum simulation
is robust. Furthermore, $U\Lambda^{-1}U\dg$ is a constant matrix that
is independent of the system parameters, which indicates that at high
temperature the quantum simulation is completely insensitive to the
nominal values of the underlying parameters.

\section{Computational aspects for the 1D transverse field Ising model}
The 1D transverse field Ising model (1D-TFIM) has a well-known mapping to a free-fermion system \cite{Lieb:1961fr,Pfeuty:1970wq}, and thus is efficiently solvable. We use these efficient solutions in order to present results for large $n$ versions of this model. In this section we explicitly demonstrate how the free fermion mapping can be used to calculate the probability distribution of the observables examined in the main text for this model. In the following we present calculations for the open boundary condition case for this model, but similar results hold for the periodic boundary condition also.

\subsection{Net magnetization distribution for the 1D-TFIM}
Recall that the Hamiltonian for the 1D-TFIM is given by
\begin{equation}
  \label{eq:111}
H=\sum_{k=1}^n B_k \sigma^k_z+ 
\sum_{j=1}^{n-1} J_{j}\sigma_x^{j} \sigma_x^{j-1}.
\end{equation}
Consider the observable $S_z=\sum_{j=1}^n \sigma_z^j = \sum_m \theta_m P_m$, where in the second equality we have decomposed the observable as a sum of projectors. We wish to compute $p_m=\Tr(P_m\varrho)$, and we use a two-step procedure to calculate this quantity. First, we express each $P_m$ as a linear combination of $\{S_1, \cdots, S_n\}$:
\begin{equation}
  \label{eq:127}
P_m=\sum_{j=1}^n \xi_{mj}S_j,
\end{equation}
where
\begin{equation}
  \label{eq:126}
  \begin{aligned}
S_1&=\sum_{k_1=1}^n \sigma_z^{k_1},\\    
S_2&=\sum_{1\le k_1\le k_2\le n}^n \sigma_z^{k_1} \sigma_z^{k_2},\\    
S_3&=\sum_{1\le k_1\le k_2\le k_3\le n}^n \sigma_z^{k_1}
\sigma_z^{k_2}\sigma_z^{k_3},\\
&\ \vdots \\
S_n&=\sigma_z^1\sigma_z^2\cdots\sigma_z^{n-1}\sigma_z^n.    
  \end{aligned}
\end{equation}
Second, we calculate the expection values of $S_j$, \ie $\la
S_j\ra=\Tr (S_j \varrho)$. Combining these two steps, we have
\begin{equation}
  \label{eq:129}
  p_m=\sum_{j=1}^n \xi_{mj}\la S_j\ra.
\end{equation}
We now elaborate on the details of these two steps. First, we express
$P_m$ in terms of $S_j$. The observable $P_m$ can be written as
\begin{equation}
  \label{eq:130}
P_m=\sum_{j=1}^{N_m} |\kappa_j\ra \la \kappa_j|,
\end{equation}
where $|\kappa_j\ra$ is a state with $m-1$ spins in the ground state
$|0\ra$ and $n-m+1$ spins in the excited state $|1\ra$, and $N_m={n
  \choose m-1}$. For simplicity, we use the case $m=2$ to
illustrate the approach. In this case, we have
\begin{equation}
  \label{eq:131}
  \begin{aligned}
P_2=& |01\cdots 1\ra \la 01\cdots 1| +|101\cdots 1 \ra \la 101\cdots
1 | +\cdots +|1\cdots 10 \ra \la 1\cdots 10 |    \\
=&|0\ra\la 0|\otimes |1\ra\la 1| \otimes \cdots \otimes |1\ra\la 1|
+|1\ra\la 1|\otimes |0\ra\la 0|\otimes |1\ra\la 1|\otimes\cdots \otimes|1\ra\la 1|
+\cdots+|1\ra\la 1|\otimes\cdots\otimes |1\ra\la 1|\otimes |0\ra\la 0|.
  \end{aligned}
\end{equation}
Since $|0\ra\la 0|={I/2}+\sigma_z$ and $|1\ra\la 1|={I/2}-\sigma_z$, we have
\begin{equation}
  \label{eq:133}
  \begin{aligned}
P_2=& \left({I/2}+\sigma_z\right)\otimes
\left({I/2}-\sigma_z\right)\otimes\cdots\otimes
\left({I/2}-\sigma_z\right)\\
& +\left({I/2}-\sigma_z\right)\otimes
\left({I/2}+\sigma_z\right)\otimes
\left({I/2}-\sigma_z\right)\otimes\cdots\otimes
\left({I/2}-\sigma_z\right)\\
&+\ \cdots\\
& +\left({I/2}-\sigma_z\right)\otimes\cdots\otimes
\left({I/2}-\sigma_z\right)\otimes
\left({I/2}+\sigma_z\right).
  \end{aligned}
\end{equation}
\Eq{eq:133} can be rewritten as 
\begin{equation}
  \label{eq:134}
  \begin{aligned}
P_2&= \left({\In/2}+\sigma_z^1\right)
\left({\In/2}-\sigma_z^2\right)\cdots
\left({\In/2}-\sigma_z^n\right)\\
& +\left({\In/2}-\sigma_z^1\right)
\left({\In/2}+\sigma_z^2\right)
\left({\In/2}-\sigma_z^3\right)\cdots
\left({\In/2}-\sigma_z^n\right)\\
&\quad \cdots\\
& +\left({\In/2}-\sigma_z^1\right)\cdots
\left({\In/2}-\sigma_z^{n-1}\right)
\left({\In/2}+\sigma_z^n\right). 
  \end{aligned}
\end{equation}
To find the coefficients $\xi_{mj}$, we replace ${\In/2}$ by
$\frac12$ and $\sigma_z^j$ by a scalar variable $x_j$ in \Eq{eq:134}
and obtain the following polynomial:
\begin{equation}
  \label{eq:8}
\begin{aligned}
p_2(x_1, \cdots, x_n)=&\left(\frac12+x_1\right)\left(\frac12-x_2\right)\cdots\left(\frac12-x_n\right)\\
&+\left(\frac12-x_1\right)\left(\frac12+x_2\right)\left(\frac12-x_3\right)\cdots\left(\frac12-x_n\right)\\
&\cdots\\
&+\left(\frac12-x_1\right)\cdots\left(\frac12-x_{n-1}\right)\left(\frac12+x_n\right).    
  \end{aligned}
\end{equation}
The polynomial $p_2$ is symmetric and thus can be represented by
elementary symmetric polynomials $s_j$:
\begin{equation}
  \label{eq:135}
  \begin{aligned}
s_1&=\sum_{k_1=1}^n x_{k_1},\\    
s_2&=\sum_{1\le k_1\le k_2\le n}^n x_{k_1} x_{k_2},\\    
s_3&=\sum_{1\le k_1\le k_2\le k_3\le n}^n x_{k_1}x_{k_2}x_{k_3},\\
&\cdots \\
s_n&=x_1x_2\cdots x_{n-1}x_n.    
  \end{aligned}
\end{equation}
The coefficients to represent $P_2$ in terms of $S_j$ are
identical to those that represent $p_2$ in terms of $s_j$, that is,
\begin{equation}
  \label{eq:136}
  p_2=\sum_{j=1}^n \xi_{2j} s_j.  
\end{equation}
In fact, to obtain $\xi_{mj}$, we can choose all the variables $x_j$ to be
the same $x$. Then, we have
\begin{equation}
  \label{eq:138}
{n \choose m} \left(\frac12+x\right)^m \left(\frac12-x\right)^{n-m}=\xi_{mn} x^n 
+{n \choose n-1} \xi_{m (n-1)} x^{n-1}
+\cdots + {n \choose 1} \xi_{m1} x+ \xi_{m0}.
\end{equation}
Equating the coefficients in both sides of \Eq{eq:138}, we can obtain $\xi_{mj}$.

Next we show how to compute $\la S_j\ra$.  From Refs. \cite{Lieb:1961fr, Pfeuty:1970wq}, we define two matrices $P$ and $Q$ as
\begin{equation}
\label{eq:3}
\begin{aligned}
P_{jk}=& \begin{cases}
 J_{j}, & \text{if } k=j+1; \\
 J_{k}, & \text{if } j=k+1; \\
B_{j}, & \text{if } j=k; \\
0, & \text{otherwise},
  \end{cases}\\
Q_{jk}=& \begin{cases}
 J_{j}, & \text{if } k=j+1;\\
-J_{k}, & \text{if } j=k+1;\\
0, & \text{otherwise}.
  \end{cases}    
  \end{aligned}
\end{equation}
Let $\phi_k^T$ be a normalized row eigenvector of $(P-Q)(P+Q)$, \ie
$\phi_k^T (P-Q)(P+Q)=\Lambda_k^2 \phi_k^T$. Let
$\psi_k^T=-\Lambda_k^{-1}\phi_k^T(P-Q)$. Juxtapose $\phi_k^T$ and
$\psi_k^T$ into two matrices $\Phi$ and $\Psi$. For the calculation of
ground state, we define
\begin{equation}
\label{eq:Gg}
G^g=\Psi^T\Phi;
\end{equation}
and for the thermal state,
we let 
\begin{equation}
\label{eq:Gt}
G^t=\Psi^T \tanh(\frac{\beta}2 \Lambda) \Phi.
\end{equation}
From Wick's
theorem and Ref.~\cite{Lieb:1961fr}, we know that
$\la S_j\ra$ is the sum of all the $j$-by-$j$ principle minor of $G$.
Moreover, from Ref.~\cite{Horn:85}, we have 
\begin{equation}
  \label{eq:137}
\det (tI-G)=t^n-\la S_1\ra t^{n-1} +\la S_2\ra t^{n-2}- \cdots
\pm \la S_n\ra.
\end{equation}
Hence we can determine $\la S_j\ra$ by calculating the
characteristic polynomial of $G$. With these two steps, we can now
obtain $p_m$. 

\subsection{Correlation function distribution for the 1D-TFIM}
When the observable is the correlation function $C_z(i,j)=\sigma_z^i
\sigma_z^j$, we know from Eq. (2.33c) in Ref.~\cite{Lieb:1961fr} that under the ground state,
\begin{equation}
\la \sigma_z^i \sigma_z^j \ra=(G_{ii}^g G_{jj}^g-G_{ji}^gG_{ij}^g)/4;
\end{equation}
and under the thermal state,
\begin{equation}
\la \sigma_z^i \sigma_z^j \ra=(G_{ii}^t G_{jj}^t-G_{ji}^tG_{ij}^t)/4,
\end{equation}
where $G^g$ and $G^t$ are defined in \Eqs{eq:Gg} and~\eqref{eq:Gt}, respectively.

We then consider to analytically calculate the FIM for ground state. Since
$\sigma_z^i\sigma_z^j$ has two eigenvalues $\pm \frac14$, we obtain
that for ground state,
\begin{equation}
  \begin{aligned}
p_1&=\tr P_1 |\psi_g\ra \la \psi_g|= (1+G_{ii}^gG_{jj}^g-G_{ji}^gG_{ij}^g)/2,\\
p_2&=\tr P_2 |\psi_g\ra \la \psi_g|= (1-G_{ii}^gG_{jj}^g+G_{ji}^gG_{ij}^g)/2.
  \end{aligned}
\end{equation}
Then
\begin{equation}
\frac{d p_1}{d \lambda_l}=\frac12\left(
\frac{d G_{ii}^g}{d\lambda_l} G_{jj}^g+
G_{ii}^g \frac{d G_{jj}^g}{d\lambda_l} 
- \frac{d G_{ji}^g}{d\lambda_l} G_{ij}^g
-G_{ji}^g \frac{d G_{ij}^g}{d\lambda_l}\right)
\quad \text{ and}  \quad 
\frac{d p_2}{d \lambda_l}=-\frac{d p_1}{d \lambda_l}.
\end{equation}
We now derive $dG^g/d\lambda_l$. Since $G^g=\Psi^T \Phi$, we have
\begin{equation}
 \frac{dG^g}{d\lambda_l}= \frac{d\Psi^T}{d\lambda_l}\Phi+
\Psi^T \frac{d\Phi}{d\lambda_l}.
\end{equation}
The matrix $(P-Q)(P+Q)$ is simple, meaning that it has pairwise
distinct eigenvalues. Then its eigenvalue and the associated
eigenvector are infinitely differentiable in a neighborhood of $\Ho$
and their differentials are
\begin{equation}
\frac{d \phi_k}{d \lambda_l}=  
\left(\Lambda_k^2 I_n-(P-Q)(P+Q)\right)^+ \left(
  \frac{d}{d\lambda_l}(P-Q)(P+Q)\right) \phi_k,
\end{equation}
where $^+$ denotes the Moore-Penrose pseudoinverse.
From the definition of $P$ and $Q$ in \Eq{eq:3}, it is straightforward
to derive $dP/d\lambda_l$ and $dQ/d\lambda_l$ and thus
\begin{equation}
\frac{d}{d\lambda_l}(P-Q)(P+Q)=
\left( \frac{dP}{d\lambda_l}-\frac{dQ}{d\lambda_l}\right)(P+Q)
+(P-Q)\left(\frac{dP}{d\lambda_l}+\frac{dQ}{d\lambda_l}\right).
\end{equation}
Moreover, we have that
\begin{equation}
  \begin{aligned}
\frac{d\psi_k^T}{d\lambda_l}&=-\frac{d \Lambda_k^{-1}}{d\lambda_l}
\phi_k^T (P-Q)-\Lambda_k^{-1} \frac{d\phi_k^T}{d\lambda_l}(P-Q)
-\Lambda_k^{-1}\phi_k^T\left(\frac{dP}{d\lambda_l}
-\frac{dQ}{d\lambda_l}\right)      \\
&=-\Lambda_k^{-1} \frac{d \Lambda_k}{d\lambda_l}\psi_k^T
-\Lambda_k^{-1} \frac{d\phi_k^T}{d\lambda_l}(P-Q)
-\Lambda_k^{-1}\phi_k^T\left(\frac{dP}{d\lambda_l}
-\frac{dQ}{d\lambda_l}\right),      
  \end{aligned}
\end{equation}
where
\begin{equation}
\frac{d \Lambda_k}{d \lambda_l}=\frac{1}{2\Lambda_k} 
\phi_k^T \left( \frac{d}{d\lambda_l}(P-Q)(P+Q)\right) \phi_k.
\end{equation}
Combining these equations, we can calculate $dp_1/d\lambda_l$ and
$dp_2/d\lambda_l$ for ground state analytically. For thermal states,
we just need to calculate an additional derivative of
$\tanh(\frac{\beta}2 \Lambda)$ in $G^t$ and can obtain the results
similarly.

\begin{figure*}[tb]
\centering
  \subfigure[Symmetry of $3\times 3$ lattice]{
     \includegraphics[scale=0.24]{3x3} }\quad\quad\quad
 \subfigure[Eigenvalues of FIM]{ \includegraphics[scale=0.23]{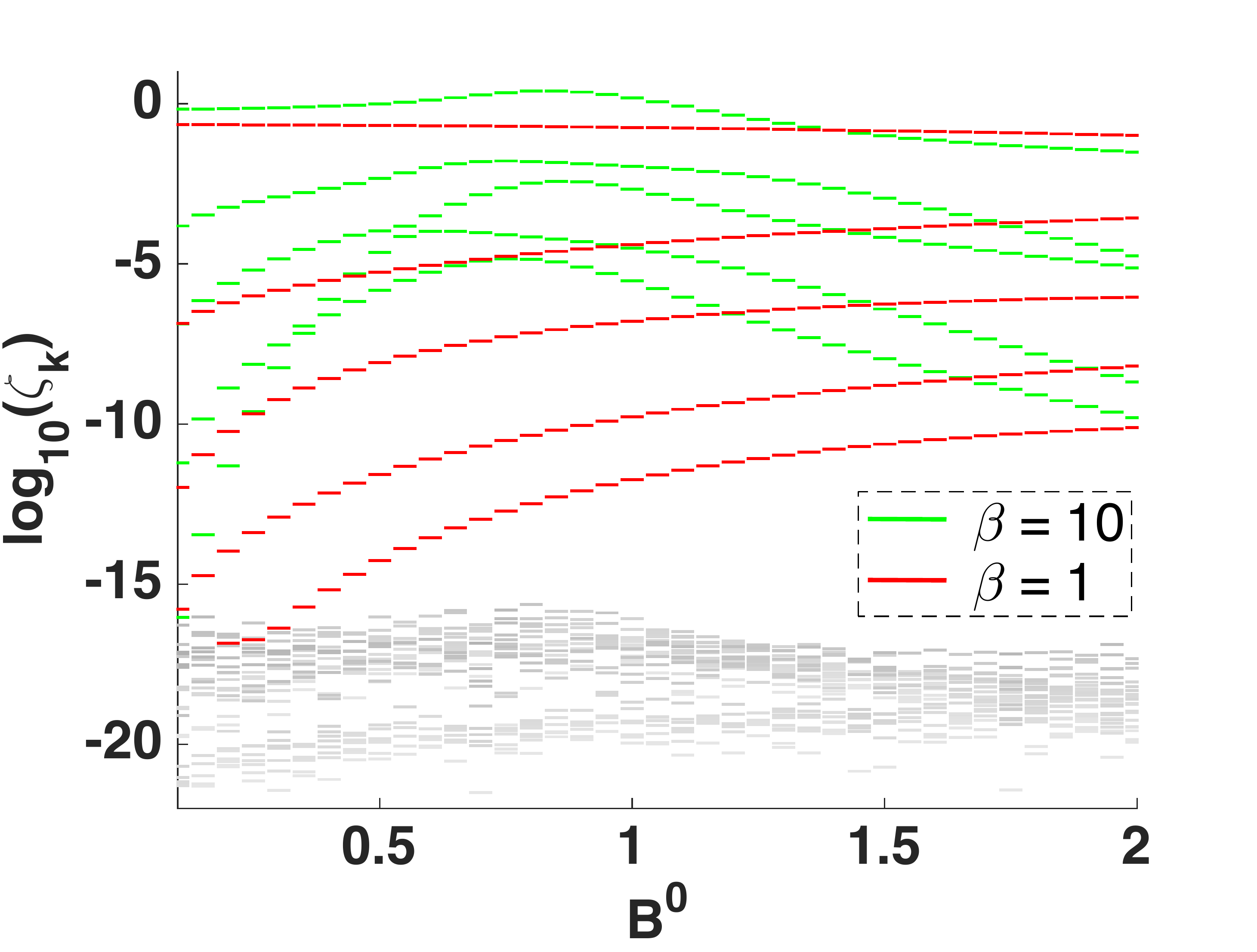} }
\\
 \subfigure[Influential CPD when $\beta=10$]{
     \includegraphics[scale=0.23]{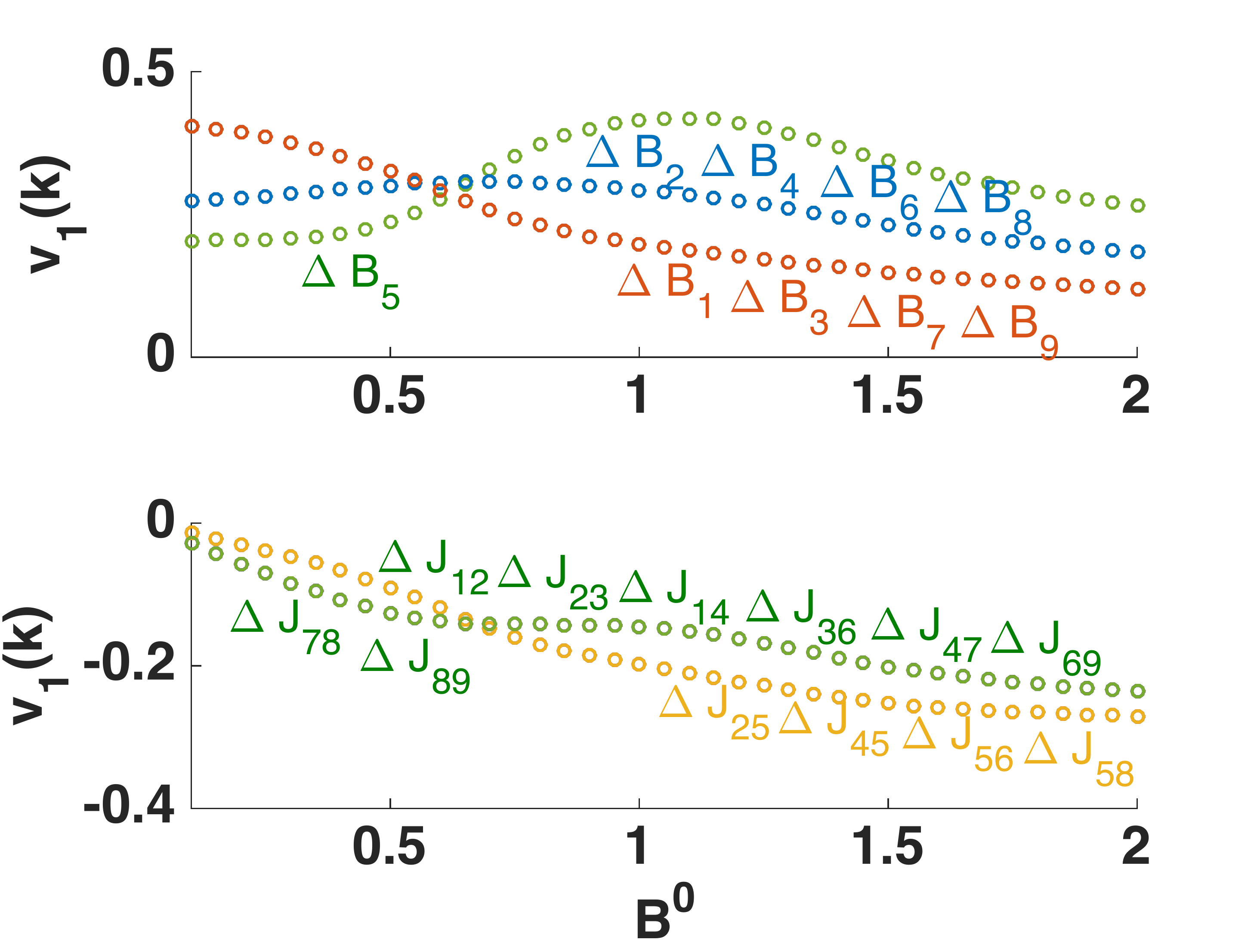} }
 \subfigure[Influential CPD when $\beta=1$]{
     \includegraphics[scale=0.21]{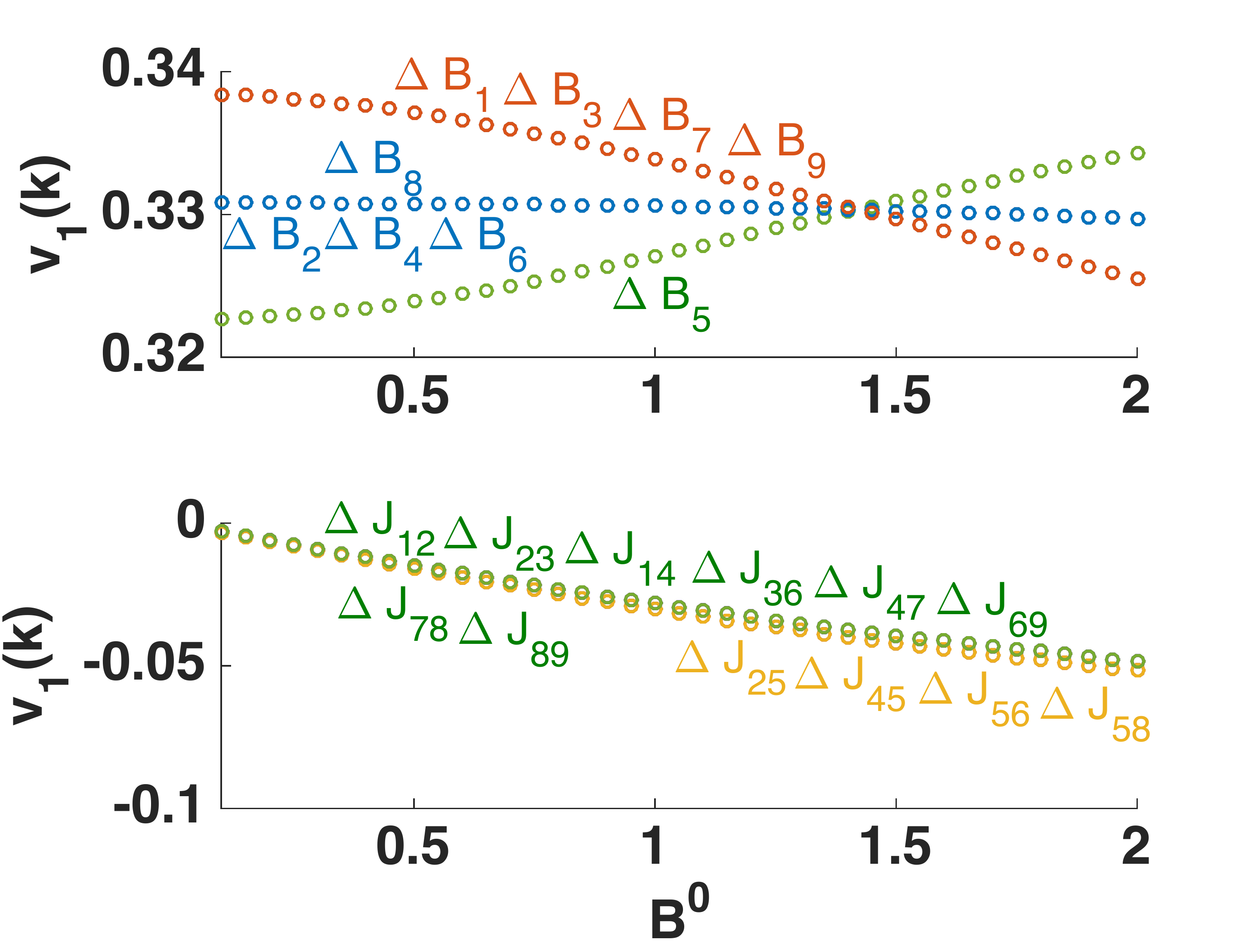} }
\caption{Eigenvalues of the FIM for the AQS model
  $\{H_2, S_z\}$, evaluated for a $3\times 3$ lattice of spins for thermal states with $\beta=10$, $1$. (a) shows the symmetry of the model, reproduced from Fig. 4 in the main text. Lattice sites and couplings that lie in the same orbit (under the reflections and rotations that leave the quantum simulation model unchanged) are identically colored. (b) shows eigenvalues of the FIM, with the five eigenvalues of largest magnitude shown in color. (c), (d) show the forms of the influential CPDs for $\beta=10$, $1$, respectively.}
\label{fig:2dIsing}
\end{figure*}

When the observables are $\sigma_x^i\sigma_x^j$ and
$\sigma_y^i\sigma_y^j$, their mean values can be obtained from
Eq. (2.33a) and (2.33b) in Ref.~\cite{Lieb:1961fr}. And following similar procedures as above, we can derive analytical expressions for derivatives of the measurement probabilities.

\section{Robustness of more quantum simulation models}
In this section we report the behavior of the FIM for some quantum simulation models that were not included in the main text for conciseness.

\subsection{2D transverse field Ising model}
In the main text we demonstrate how symmetry analysis of the 2D-TFIM with open boundary conditions and net magnetization as the observable enables one to determine the rank of FIM for this model, and show that it is sloppy. For more details on the symmetry analysis for this model, see section \ref{sec:sym_rep} in this Appendix. Here in Fig. \ref{fig:2dIsing}, we explicitly present the eigenvalues and eigenvectors of the FIM for a $3\times 3$ square lattice version of this model. It is evident from Fig. \ref{fig:2dIsing}(b) that the FIM eigenvalues agree with the rank bound (rank $\leq 5$) derived from symmetry. Furthermore, Figs. \ref{fig:2dIsing}(c) and (d) show that the forms of the influential CPDs respect the symmetry of the model. 

\subsection{1D random Ising model}
\begin{figure}[tb]
\centering
 \subfigure[Eigenvalues of FIM]{ \includegraphics[scale=0.24]{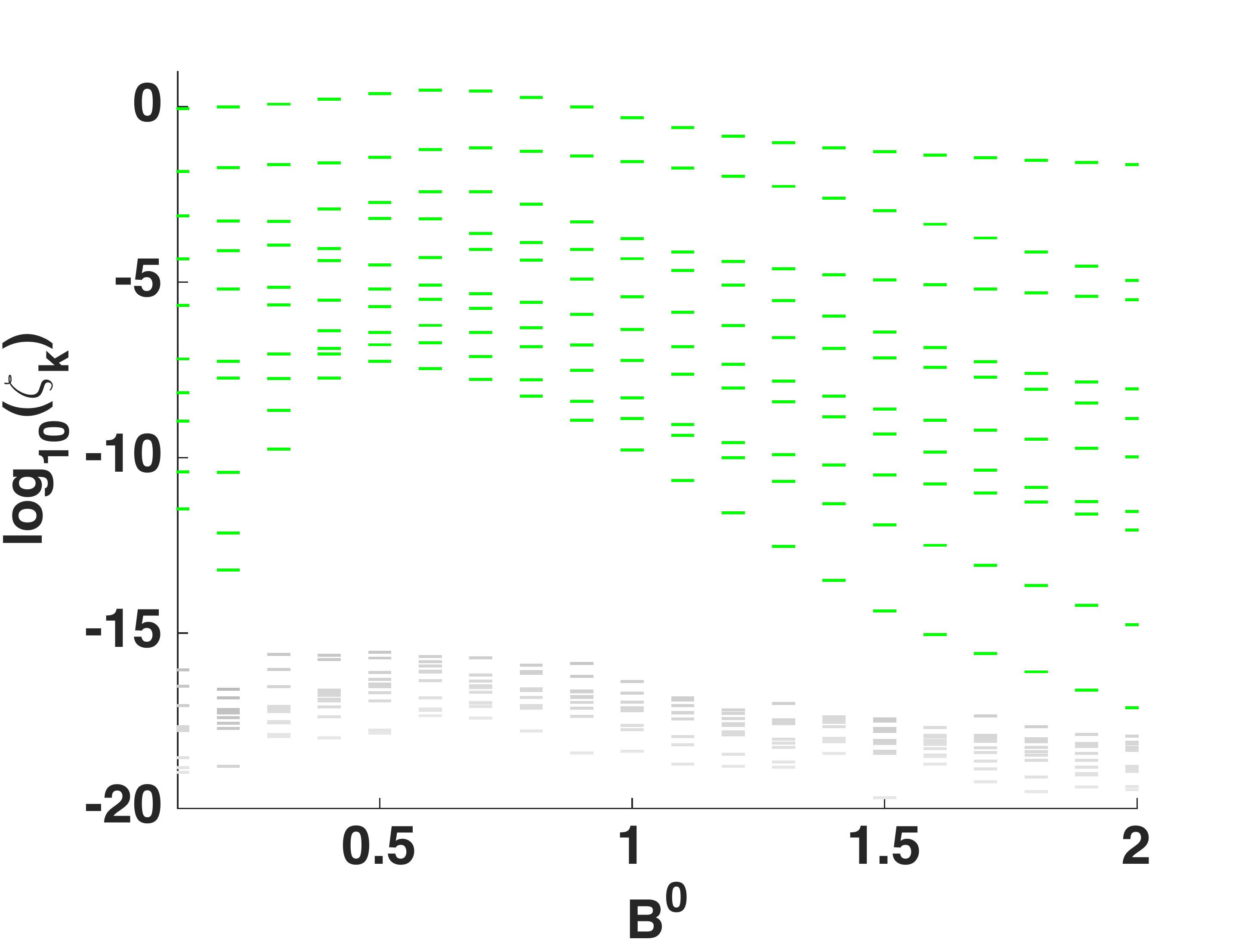} }
\subfigure[Influential CPD]{
     \includegraphics[scale=0.24]{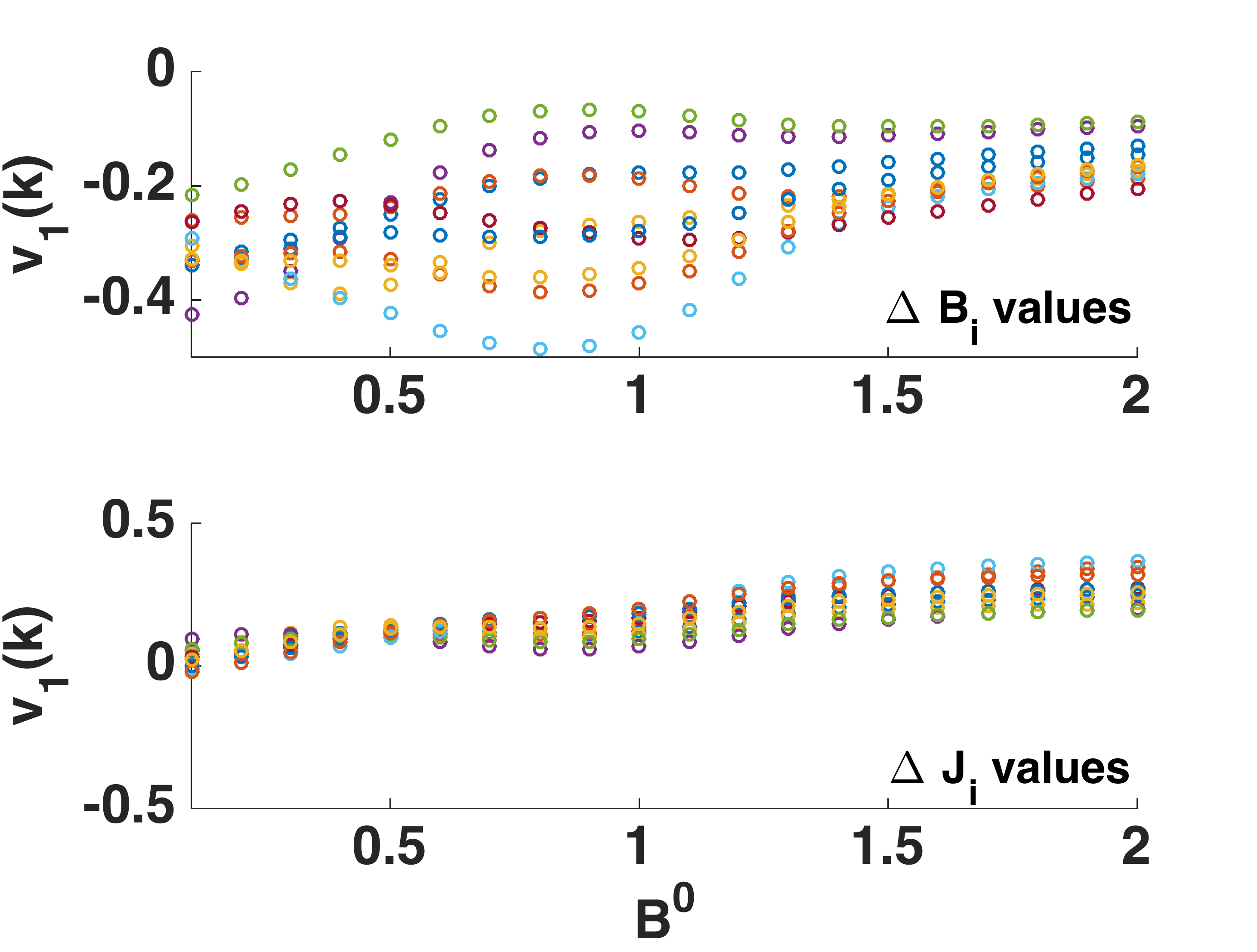} }
\caption{(a) Eigenvalues of the FIM for the AQS model $\{H_1^R, S_z\}$, for a $10$-spin model under the thermal state with $\beta=10$. A bound derived from considering the number of measurement outcomes tells us that the rank of the FIM is at most $10$, and therefore we show the ten largest eigenvalues in color and the others (numerical artifacts) in gray. (b) The form of the first influential CPD. }
\label{fig:randomIsing}
\end{figure}

To examine a model with disorder, consider the 1D transverse field Ising model with random local fields and coupling energies, \ie
\beq
H_1^R = \sum_{i=1}^n B_i^0 \sigma_z^i + \sum_{i} J_{i}^0 \sigma_x^i \sigma_x^{i+1},
\eeq
with periodic boundary conditions ($\sigma_x^{n+1}\equiv \sigma_x^1$), and $B_i^0 = B^0 + \delta B_i, J_i^0 = J^0 + \delta J_i$, where $\delta B_i$ and $\delta J_i$ are independent zero-mean Gaussian random variables with standard deviation $\sigma$. As for the observable of interest, consider the net magnetization $S_z$ again. This quantum simulation model has no symmetries due to the random parameters and so the FIM rank bounds based on symmetry are not informative. The number of measurement outcomes for this observable is $M=n+1$, and therefore the rank of the FIM is at most $n$. In Fig. \ref{fig:randomIsing}(a) we show the eigenvalues of the FIM for a $10$-spin example of this quantum simulation model, with $J^0=1$, disorder variance $\sigma=0.2$ and $\beta=10$. This figure shows the FIM eigenvalues for one representative sample of $\delta B_i$ and $\delta J_i$. As evident from this figure, while the dominant eigenvalue is roughly two orders of magnitude above all others, this model cannot be considered sloppy except for small or large values of $B^0$. In Fig. \ref{fig:randomIsing}(b) we also show the form of the first influential CPD (we do not label the points on this plot since we only wish to illustrate the complexity of the behavior of this quantity for this model). 

\subsection{$J_1$-$J_2$ antiferromagnetic Heisenberg model}
\begin{figure}[h]
\centering
     \includegraphics[scale=0.24]{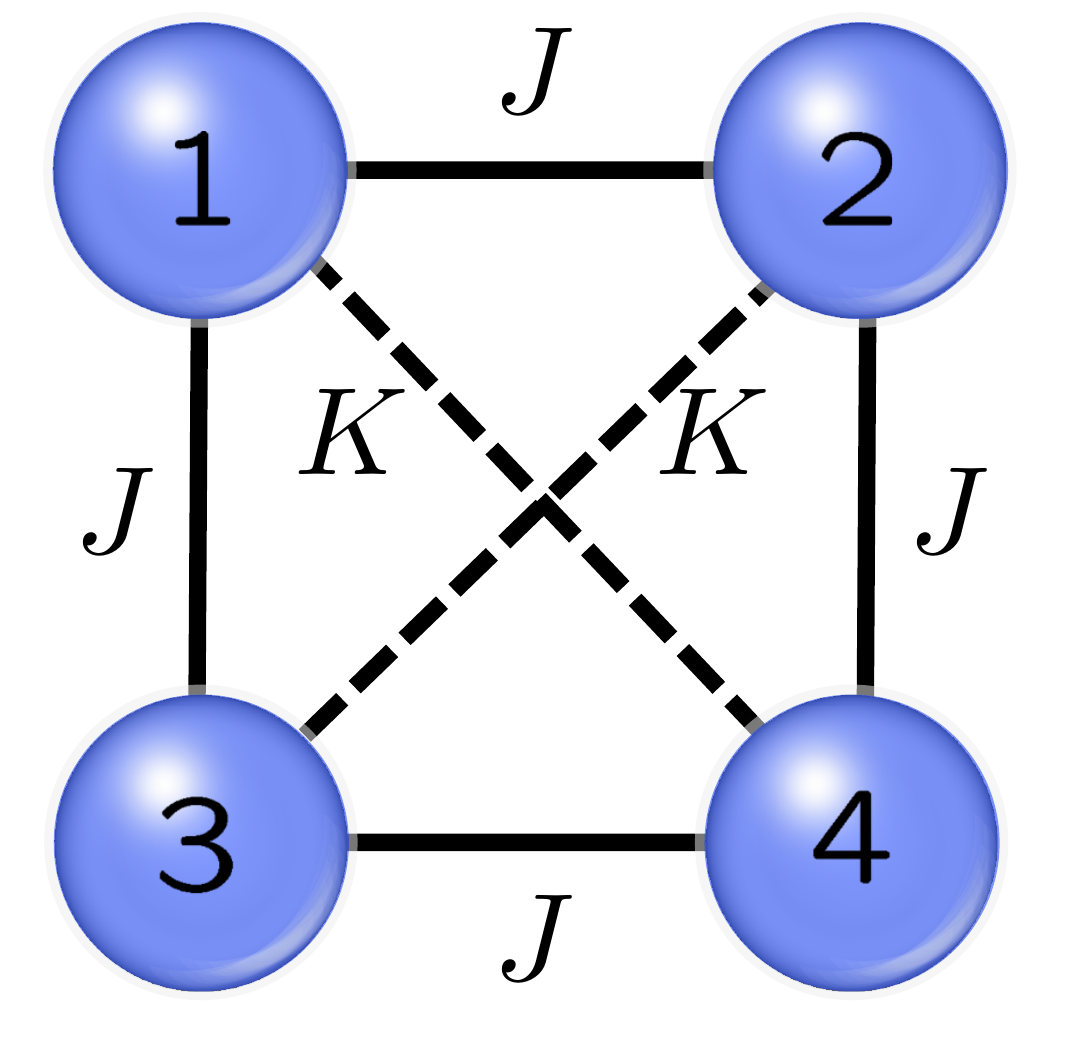} 
\caption{A single plaquette defining interactions between spins for a $J_1$-$J_2$ Heisenberg model.}
\label{fig:J1J2_plaquette}
\end{figure}

Now we turn to a quantum simulation model based on a Hamiltonian that contains non-nearest-neighbor interactions and geometric frustration. The $J_1$-$J_2$ antiferromagnetic Heisenberg model is defined by the following Hamiltonian governing spin-$1/2$ systems on a two-dimensional lattice:
\beq
H_5 = \sum_{\langle i,j\rangle} J_{ij} \sigma^i \cdot \sigma^j + \sum_{\langle\langle i,j \rangle \rangle} K_{ij} \sigma^i \cdot \sigma^j ,
\eeq
where the first sum is over nearest-neighbor spins and the second is over next-nearest-neighbor spins. We are interested in the uniform nominal operating point for this model where $J_{ij}^0 = J^0$ and $K_{ij}^0 = K^0$ with $J^0$, $K^0>0$ \footnote{Conventionally the parameters in this model are $J_1$ (instead of $J^0$) and $J_2$ (instead of $K^0$), and hence the name for the model. However, to simplify notation, we use the above parameter names.}. Fig.~\ref{fig:J1J2_plaquette} shows a single plaquette in the square lattice in the nominal model. 

The magnetic order in this system is complex with different phases of magnetic ordering being driven by competition between the two different kinds of interactions. The magnetic order parameter is different in different $K^0/J^0$ regimes. For small values of this ratio ($\sim 0$) the magnetization is N\'eel ordered (the model resembles a conventional Heisenberg antiferromagnet on a square lattice in this regime), and as this ratio approached unity one has so-called ``striped magnetization'' \cite{Morita:2015dz}. Our observables of interest is the staggered magnetization, which probes the N\'eel order in the system:
\beq
M_s = \sum_{j=1}^{n} \sum_{i<j} (-1)^{j-i} \sigma^i \cdot \sigma^j,
\eeq
where $n$ is the total number of spins in the system.

\begin{figure*}[t!]
\centering
  \subfigure[Symmetry of $3\times 3$ lattice]{
     \includegraphics[scale=0.2]{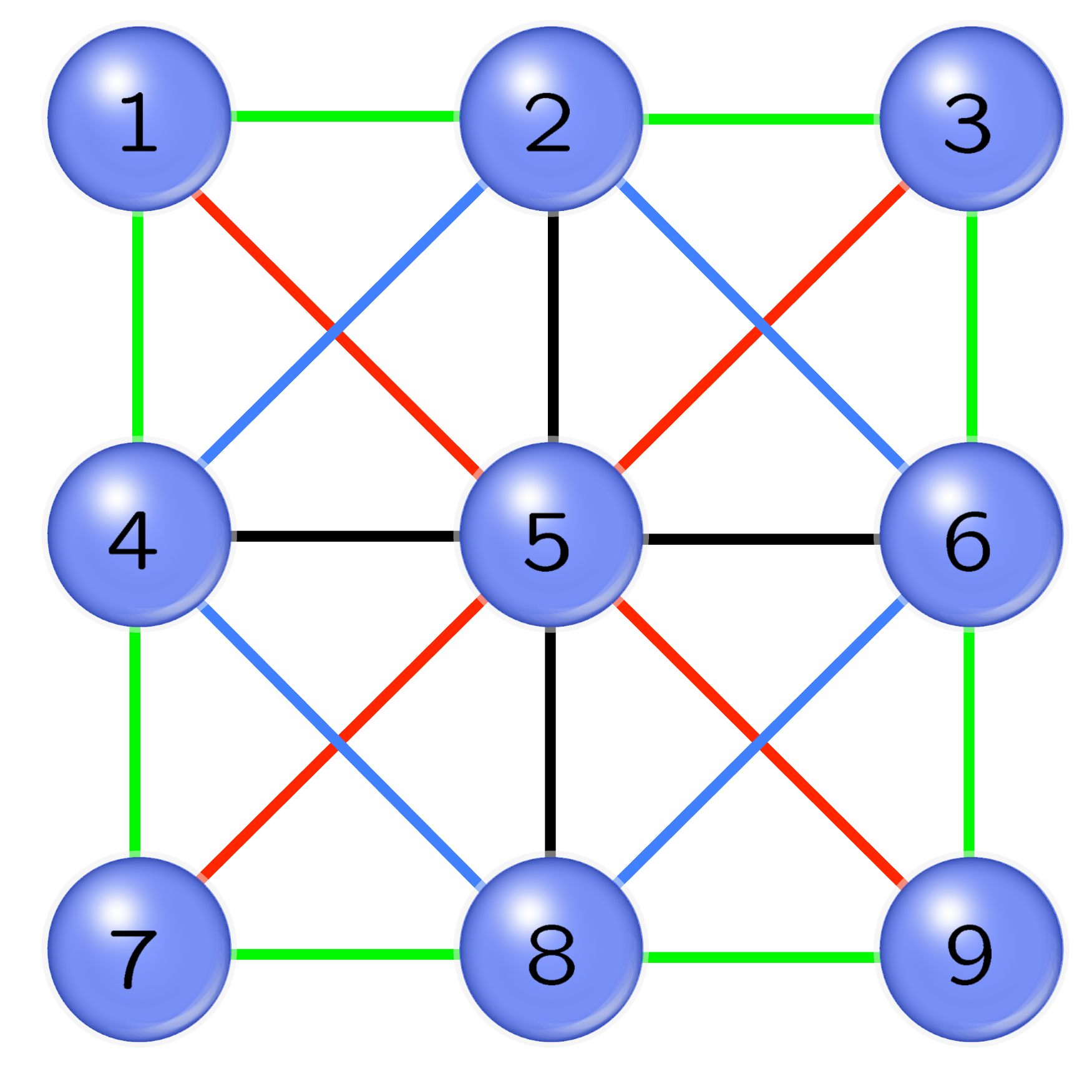} }\quad\quad\quad
 \subfigure[Eigenvalues of FIM]{ \includegraphics[scale=0.2]{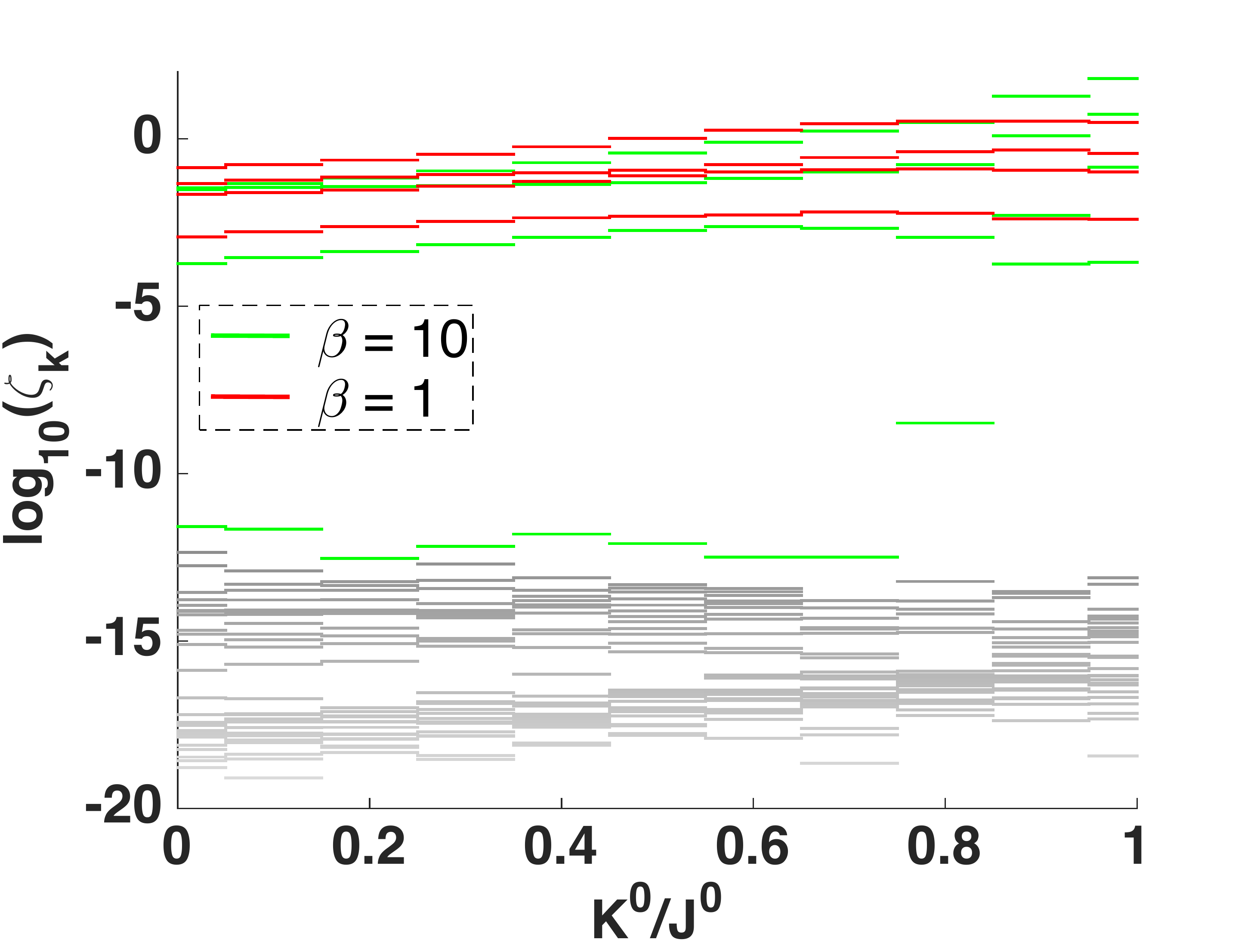} }
 \subfigure[Influential CPD when $\beta=10$]{
     \includegraphics[scale=0.2]{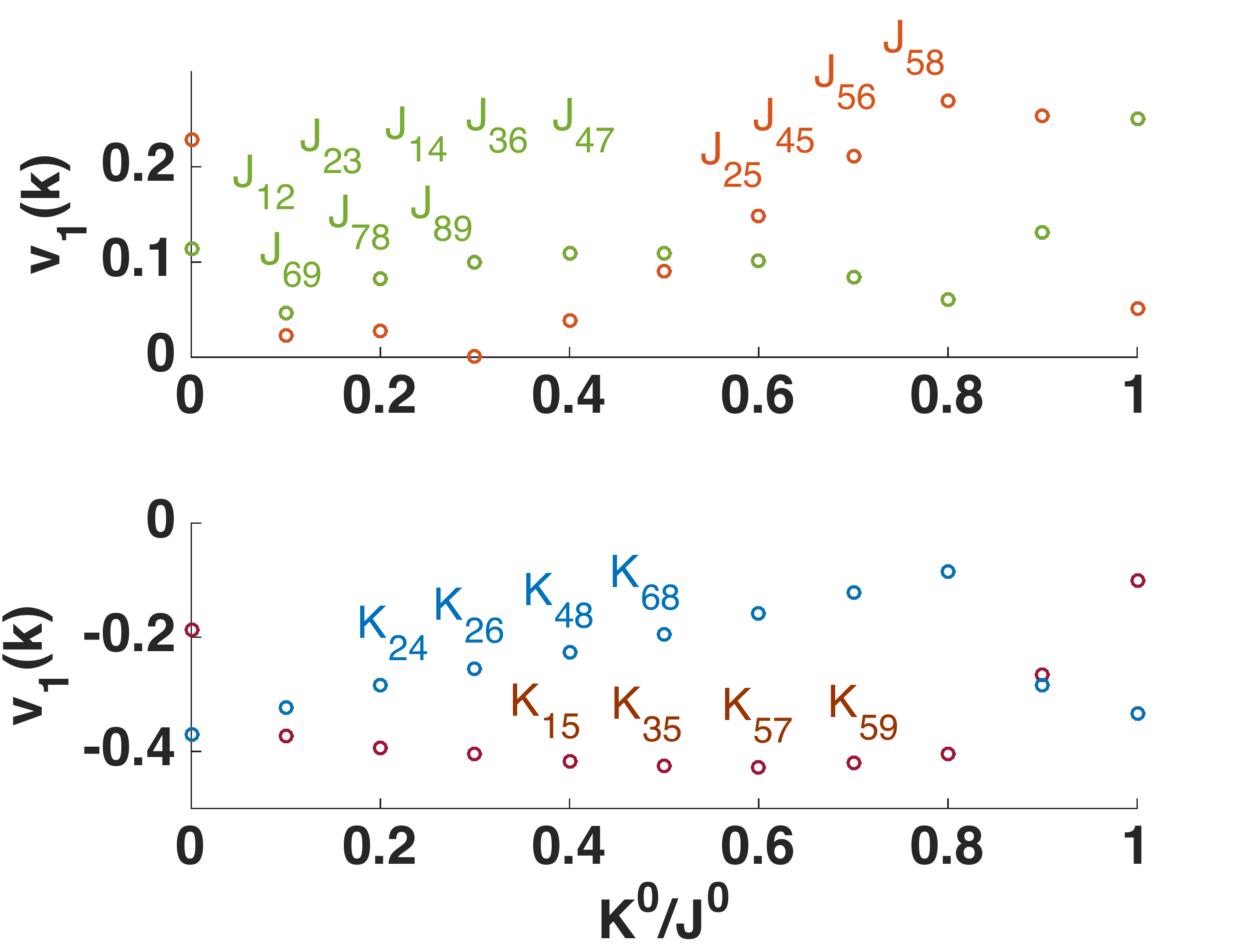} }
\caption{Eigenvalues of the FIM for the AQS model
  $\{H_5, M_s\}$, evaluated for a $3\times 3$ lattice of spins for thermal states with $\beta=1$, $10$. (a) shows the symmetry of the model. Couplings that lie in the same orbit (under the reflections and rotations that leave the quantum simulation model unchanged) are identically colored. (b) shows eigenvalues of the FIM, with the five eigenvalues of largest magnitude shown in color. (c) shows the form of the influential CPD for $\beta=10$.}
\label{fig:J1J2}
\end{figure*}

The quantum simulation model $\{H_5, M_s\}$ with open boundary conditions on the lattice has several symmetries despite the complicated form of the observable of interest. For square lattices, this model has rotational symmetry about the center of the lattice and reflection symmetry about four reflection lines. In Fig. \ref{fig:J1J2}(a) we explicitly show the symmetries in this model for a $3\times 3$ square lattice. Note that since $n$ is odd, all these symmetry transformations take odd (even) labeled spins to odd (even) labelled spins, and hence leave the observable of interest invariant. From this symmetry analysis, we obtain a rank bound on the FIM of $\rk(F)\leq 4$. Fig.  \ref{fig:J1J2}(b) shows the eigenvalues of the FIM for this $3\times 3$ example for $\beta=10$ and $\beta=1$, and it is clear that the rank bound is respected. Finally, Fig. \ref{fig:J1J2}(c) shows the primary influential CPD for this model when $\beta=10$. The first four eigenvectors of the FIM all define influential CPDs since the first four eigenvalues are non-negligible. We only plot the primary influential CPD here for simplicity, but all the others have the same symmetry properties. 

\section{Examples of model symmetries and representations}
\label{sec:sym_rep}
Here we explicitly construct representations of symmetry groups for two quantum simulation models analyzed in the main text. These representations acting on the Hilbert space of the model can be constructed from elementary SWAP operations.
 
First consider the 1D transverse-field Ising model (1D-TFIM) with periodic boundary conditions
$\{\Ha^{\rm per}, S_z\}$ as discussed in the main text.
\begin{center}
\includegraphics[scale=0.2]{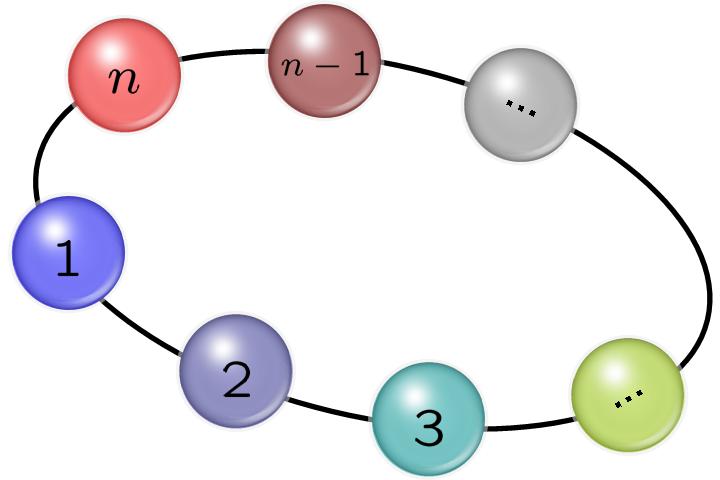} 
\end{center}
This model is translationally invariant and therefore its symmetry group $G$ is defined as
\begin{equation*}
  G=\{I, U, U^2, \cdots, U^{n-1}\},
\end{equation*}
where
\begin{equation*}
 U=U_{n-1,n}\cdots U_{23} U_{12},
\end{equation*}
and $U_{jk}$ is the SWAP operation between two nodes $j$ and $k$, \ie
\begin{equation*}
U_{jk}=2\sigma_x^j\sigma_x^k+2\sigma_y^j\sigma_y^k + 2\sigma_z^j
  \sigma_z^k +I/2.
\end{equation*}
It is easy to verify that 
\begin{equation*}
  U_{jk} \sigma_{\w}^m U_{jk}^\dag=
  \begin{cases}
\sigma_{\w}^k, &\text{ if } m=j; \\
\sigma_{\w}^j, &\text{ if } m=k; \\
\sigma_{\w}^m,& \text{ otherwise},\\
  \end{cases}
\end{equation*}
where ${\w}=x$, $y$, or $z$. For $m<n$, we have
\begin{equation*}
  U_{jk} \sigma_{\w}^m \sigma_{\w}^n U_{jk}^\dag=
  \begin{cases}
\sigma_{\w}^m \sigma_{\w}^n, &\text{ if } m=j \text{
 and } n=k, \text{ or }  m\neq j \text{ and } n\neq k; \\
\sigma_{\w}^k \sigma_{\w}^n, &\text{ if } m=j \text{ and } n\neq k;\\
\sigma_{\w}^m \sigma_{\w}^k, &\text{ if } m\neq j \text{ and } n=k;\\
\sigma_{\w}^j \sigma_{\w}^n, &\text{ if } m=k;\\
\sigma_{\w}^m \sigma_{\w}^k, &\text{ if } n=j.
  \end{cases}
\end{equation*}
Therefore, we can obtain
\begin{equation*}
  \label{eq:5}
 U \sigma_{\w}^1 U^\dag=\sigma_{\w}^n, \quad
U \sigma_{\w}^2 U^\dag=\sigma_{\w}^1, \quad
\cdots,\quad
U \sigma_{\w}^n U^\dag=\sigma_{\w}^{n-1}, 
\end{equation*}
and
\begin{equation*}
  \label{eq:10}
    U \sigma_{\w}^1\sigma_{\w}^2 U^\dag=\sigma_{\w}^n\sigma_{\w}^1, \quad
U \sigma_{\w}^2\sigma_{\w}^3 U^\dag=\sigma_{\w}^1\sigma_{\w}^2, \quad
\cdots,\quad
U \sigma_{\w}^n\sigma_{\w}^1 U^\dag=\sigma_{\w}^{n-1}\sigma_{\w}^n.
\end{equation*}
For any $g$, we have that
\begin{equation}
  \label{eq:15}
 U^g \sigma_{\w}^j (U^g)^\dagger=\sigma_{\w}^{j-g}, \quad
U^g \sigma_{\w}^j\sigma_{\w}^{j+1} (U^g)^\dagger=  \sigma_{\w}^{j-g}\sigma_{\w}^{j-g+1}, 
\end{equation}
where $j-g$ is understood to be computed with modulo $n$.  Then, since the ideal Hamiltonian for the model has identical nominal parameters ($B_i^0 = B^0, J_i^0 = J^0$), we have that
$U^g H (U^g)^\dagger=H$; and furthermore, $U^gO(U^g)^\dagger=O$. From \Eq{eq:15}
and the discussion in \app{C}, we know that
\begin{equation*}
  \label{eq:49}
\frac{\partial p_m(\lvec)} {\partial B_j}=
\frac{\partial p_m(\lvec)} {\partial B_{k}},\quad
\frac{\partial p_m(\lvec)} {\partial J_j}=
\frac{\partial p_m(\lvec)} {\partial J_{k}}, 
\quad j, k=1, \cdots, n.
\end{equation*}
We can thus write $V$ as
\begin{equation*}
V=\ma{\Id\cdot a^T \\ \Id\cdot b^T},
\end{equation*}
where $\Id=\ma{1&\cdots&1}^T\in \R^n$, $a$, $b\in \R^n$, and the FIM
can be written as
\begin{equation*}
F=\ma{a^T\\b^T} \Lambda^{-1} \ma{a&b} \otimes ( \Id\cdot \Id^T).
\end{equation*}
From this form it is evident that $\rk F=2$, and the two nonzero eigenvectors are
\begin{equation*}
\ma{\mu&\cdots&\mu&\eta&\cdots&\eta}^T
\end{equation*}
and
\begin{equation*}
\ma{-\eta&\cdots&-\eta&\mu&\cdots&\mu}^T.
\end{equation*}
Hence the influential composite error deviations take the form $\sum_i\Delta B_i$ and
$\sum_i\Delta J_i$.
 
\medskip
 
Next consider a 2D-TFIM on a square lattice with open boundary conditions. A more explicit form of the Hamiltonian for this model than the one given in the main text is:
\begin{equation*}
H_2= \sum_{j_1=-n}^n \sum_{j_2=-n}^n B_{(j_1, j_2)} \sigma_x^{(j_1, j_2)} 
+\sum_{j_1=-n}^n \sum_{j_2=-n}^{n-1} J_{(j_1, j_2)}^{(j_1, j_2+1)} 
\sigma_z^{(j_1, j_2)} \sigma_z^{(j_1, j_2+1)} 
+\sum_{j_1=-n}^{n-1} \sum_{j_2=-n}^{n} J_{(j_1, j_2)}^{(j_1+1, j_2)} 
\sigma_z^{(j_1, j_2)} \sigma_z^{(j_1+1, j_2)},
\end{equation*}
where $(j_1,j_2)$ denotes the Cartesian coordinate for a node, \eg 

\begin{center}
\includegraphics[scale=0.5]{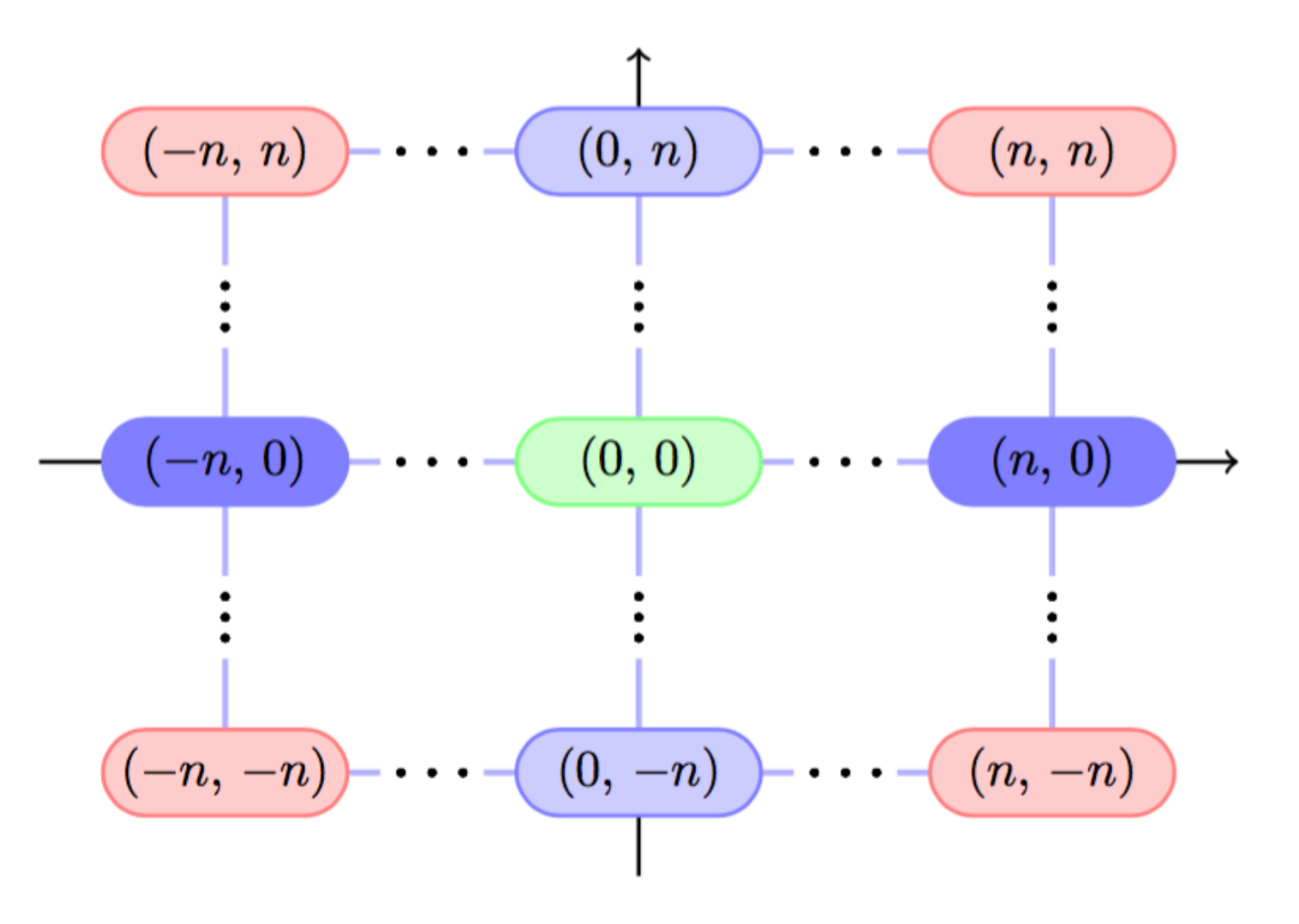} 
\end{center}
 
 The quantum simulation model we consider is $\{H_3, S_z\}$, and thus the observable has complete translational symmetry. For the nominal values of the parameters for this quantum simulation model, we only require those that are symmetric with respect to $x$- or $y$-axes are equal, \ie
\begin{eqnarray*}
\label{eq:48}
B^0_{(j_1, j_2)}&=&B^0_{(-j_1, j_2)}, \quad B^0_{(j_1, j_2)}=B^0_{(j_1, -j_2)},\\
\label{eq:91}
\left( J_{(j_1, j_2)}^{(j_1, j_2+1)}\right)^0 &=& \left(J_{(j_1, -j_2-1)}^{(j_1, -j_2)}\right)^0 , \quad 
\left(J_{(j_1, j_2)}^{(j_1+1, j_2)}\right)^0=\left(J_{(-j_1-1, j_2)}^{(-j_1, j_2)}\right)^0.
\end{eqnarray*}

In this case the quantum simulation model has reflection symmetry about the $x$- and $y$- axes and $90^\circ$ rotation symmetry. The generators of the symmetry group are $\{I, U_x, U_y, U_R\}$, where
\begin{equation*}
  \label{eq:61}
  \begin{aligned}
U_x=&\prod_{j_1=-n}^n \prod_{j_2=1}^n M_{(j_1,  j_2)}^{(j_1, -j_2)},\\
U_y=&\prod_{j_1=1}^n \prod_{j_2=-n}^n M_{(j_1,  j_2)}^{(-j_1, j_2)},\\
U_R=&\prod_{j_1=-n}^{-1} \prod_{j_2=-n}^{0} M_{(j_1,  j_2)}^{(j_2, -j_1)}
\prod_{j_1=-n}^0 \prod_{j_2=1}^n M_{(j_1,  j_2)}^{(j_2, -j_1)}
\prod_{j_1=1}^{n} \prod_{j_2=0}^n M_{(j_1,  j_2)}^{(j_2, -j_1)},
  \end{aligned}
\end{equation*}
where $M_{(j_1, j_2)}^{(k_1, k_2)}$ is the SWAP operation between two
nodes $(j_1, j_2)$ and $(k_1, k_2)$:
\begin{equation*}
M_{(j_1, j_2)}^{(k_1, k_2)}= 
2\sigma_x^{(j_1,j_2)}\sigma_x^{(k_1,k_2)}
+2\sigma_y^{(j_1,j_2)}\sigma_y^{(k_1,k_2)}
+2\sigma_z^{(j_1,j_2)}\sigma_z^{(k_1,k_2)}+I/2.
\end{equation*}
When this operator is applied to local terms, we have
\begin{equation*}
  \label{eq:60}
 M_{(j_1, j_2)}^{(k_1, k_2)} \sigma_\w^{(m_1,m_2)}
M_{(j_1, j_2)}^{\dag (k_1, k_2)} =
\begin{cases}
  \sigma_\w^{(k_1, k_2)}, & \text{if } m_1=j_1, m_2=j_2;\\
 \sigma_\w^{(j_1, j_2)}, & \text{if } m_1=k_1, m_2=k_2;\\
 \sigma_\w^{(m_1, m_2)}, & \text{otherwise}, 
\end{cases}
\end{equation*}
where $\w=x$, $y$, or $z$. Note that $U_x$ flips
$\sigma_\w^{(m_1,m_2)}$ with respect to $x$-axis, and $U_y$ flips
with respect to $y$-axis. The operator $U_R$ is the product of three
rotations from quadrant I to II, II to III, and III to IV, and then it
rotates $\sigma_\w^{(m_1,m_2)}$ by $90^\circ$ clockwise. Hence $U_x$,
$U_y$, and $U_R$ commute with both $\Ho$ and $O$. From the discussion
in \app{A}, we obtain
\begin{equation*}
\label{eq:73}
\begin{aligned}
 \frac{\partial p_m(\lvec)}{\partial B_{(j_1, j_2)}}&=
 \frac{\partial p_m(\lvec)}{\partial B_{(j_1, -j_2)}}, \quad
  \frac{\partial p_m(\lvec)}{\partial B_{(j_1, j_2)}}=
 \frac{\partial p_m(\lvec)}{\partial B_{(-j_1, j_2)}}, \\
 \frac{\partial p_m(\lvec)}{\partial B_{(j_1, j_2)}}&=
 \frac{\partial p_m(\lvec)}{\partial B_{(-j_2, j_1)}}=
  \frac{\partial p_m(\lvec)}{\partial B_{(-j_1, -j_2)}}=
 \frac{\partial p_m(\lvec)}{\partial B_{(j_2, -j_1)}}, 
  \end{aligned}
\end{equation*}
and
\begin{equation*}
  \begin{aligned}
\frac{\partial p_m(\lvec)}{\partial J_{(j_1, j_2)}^{(j_1,j_2+1)}}&= 
\frac{\partial p_m(\lvec)}{\partial J_{(j_1,-j_2-1)}^{(j_1, -j_2)}},\quad
\frac{\partial p_m(\lvec)}{\partial J_{(j_1, j_2)}^{(j_1, j_2+1)}}= 
\frac{\partial p_m(\lvec)}{\partial J_{(-j_1,j_2)}^{(-j_1, j_2+1)}},\\
\frac{\partial p_m(\lvec)}{\partial J_{(j_1, j_2)}^{(j_1,j_2+1)}}&= 
\frac{\partial p_m(\lvec)}{\partial J_{(-j_2,j_1)}^{(-j_2-1, j_1)}}=
\frac{\partial p_m(\lvec)}{\partial J_{(-j_1, -j_2)}^{(-j_1, -j_2-1)}}= 
\frac{\partial p_m(\lvec)}{\partial J_{(j_2,-j_1)}^{(j_2+1, -j_1)}},\\
\frac{\partial p_m(\lvec)}{\partial J_{(j_1, j_2)}^{(j_1+1, j_2)}}&= 
\frac{\partial p_m(\lvec)}{\partial J_{(-j_1-1, j_2)}^{(-j_1, j_2)}},\quad
\frac{\partial p_m(\lvec)}{\partial J_{(j_1, j_2)}^{(j_1+1, j_2)}}= 
\frac{\partial p_m(\lvec)}{\partial J_{(j_1,-j_2)}^{(j_1+1,-j_2)}},\\
\frac{\partial p_m(\lvec)}{\partial J_{(j_1, j_2)}^{(j_1+1, j_2)}}&= 
\frac{\partial p_m(\lvec)}{\partial J_{(-j_2, j_1)}^{(-j_2, j_1+1)}}=
\frac{\partial p_m(\lvec)}{\partial J_{(-j_1, j_2)}^{(-j_1-1, j_2)}}= 
\frac{\partial p_m(\lvec)}{\partial J_{(-j_2,-j_1)}^{(-j_2, -j_1-1)}}.    
  \end{aligned}
\end{equation*}
We know that all the nodes and couplings that are mirror images of
each other with respect to the horizontal or vertical axes, or images
of $90^\circ$, $180^\circ$, or $270^\circ$ rotations, have identical rows in the FIM.

\end{document}